\documentclass[10pt,english,twocolumn, journal]{IEEEtran}

\usepackage[T1]{fontenc}
\usepackage[latin1]{inputenc}
\usepackage{geometry}
\usepackage{amsmath,amssymb,amsthm,mathrsfs,amsfonts,dsfont}

\makeatletter
\def\set@curr@file#1{%
  \begingroup
    \escapechar\m@ne
    \xdef\@curr@file{\expandafter\string\csname #1\endcsname}%
  \endgroup
}
\def\quote@name#1{"\quote@@name#1\@gobble""}
\def\quote@@name#1"{#1\quote@@name}
\def\unquote@name#1{\quote@@name#1\@gobble"}
\makeatother

\usepackage{graphicx}
\usepackage{epstopdf}
\usepackage{cite}
\usepackage{color}
\usepackage{mathtools}
\usepackage{cases}
\usepackage{stackengine}
\usepackage{accents}
\usepackage{caption2}
\usepackage{float}
\usepackage{verbatim}
\usepackage{subfigure}
\usepackage{color}
\usepackage{multirow}
\usepackage{cancel}
\usepackage[shortlabels]{enumitem}

\newcommand\blfootnote[1]{%
\begingroup
\renewcommand\thefootnote{}\footnote{#1}%
\addtocounter{footnote}{-1}%
\endgroup
}

\newtheorem{thm}{Theorem}

\newtheorem{lem}{Lemma}

\newtheorem{rem}{Remark}

\renewcommand\appendix{\par
\setcounter{section}{0}
\setcounter{subsection}{0}
\setcounter{figure}{0}
\setcounter{table}{0}
\renewcommand\thesection{ \Alph{section}}
\renewcommand\thefigure{\Alph{section}\arabic{figure}}
\renewcommand\thetable{\Alph{section}\arabic{table}}
}

\usepackage{babel}

\allowdisplaybreaks

\begin{document}

\title{On Capacity-Achieving Distributions for Complex AWGN Channels Under Nonlinear Power Constraints and their Applications to SWIPT}


\author{Morteza Varasteh$^\dagger$, Borzoo Rassouli$^*$ and Bruno Clerckx$^\dagger$\\
$^\dagger$ Department of Electrical and Electronic Engineering, Imperial College London, UK.\\
$^*$ School of Computer Science and Electronic Engineering, University of Essex, UK.\\
\{m.varasteh12; b.clerckx\}@imperial.ac.uk, b.rassouli@essex.ac.uk.}

\maketitle

\begin{abstract}
The capacity of a complex and discrete-time memoryless \textit{additive white Gaussian noise} (AWGN) channel under three constraints, namely, input average power, input amplitude and output delivered power is studied. The output delivered power constraint is modelled as the average of linear combination of even moments of the channel input being larger than a threshold.
It is shown that the capacity of an AWGN channel under transmit average power and receiver delivered power constraints is the same as the capacity of an AWGN channel under an average power constraint. However, depending on the two constraints, the capacity can be either achieved by a Gaussian distribution or arbitrarily approached by using time-sharing between a Gaussian distribution and On-Off Keying.
As an application, a \textit{simultaneous wireless information and power transfer} (SWIPT) problem is studied, where an experimentally-validated nonlinear model of the harvester is used. It is shown that the delivered power depends on higher order moments of the channel input. Two inner bounds, one based on complex Gaussian inputs and the other based on further restricting the delivered power are obtained for the \textit{Rate-Power} (RP) region. For Gaussian inputs, the optimal inputs are zero mean and a tradeoff between transmitted information and delivered power is recognized by considering asymmetric power allocations between inphase and quadrature subchannels. Through numerical algorithms, it is observed that input distributions (obtained by restricting the delivered power) attain larger RP region compared to Gaussian input counterparts. The benefits of the newly developed and optimized input distributions are also confirmed and validated through realistic circuit simulations. The results reveal the crucial role played by the \textit{energy harvester} (EH) nonlinearity on SWIPT and provide new engineering guidelines on how to exploit this nonlinearity in the design of SWIPT modulation, signal and architecture. \blfootnote{This work has been partially supported by the EPSRC of the UK, under the grant EP/P003885/1.}\blfootnote{This work has been partially presented in the Information Theory Workshop 2017 (ITW) \cite{Varasteh_Rassouli_Clerckx_ITW_2017}.}
\end{abstract}

\section{Introduction}\label{Sec_Intro}
\textit{Radio-Frequency} (RF) waves can be utilized for transmission of both information and power simultaneously. Recent wireless network designs call for unifying wireless transmission of information and power so as to make the best use of the RF spectrum and radiation as well as the network infrastructure for the dual purpose of communicating and energizing \cite{Clerckx_Zhang_Schober_Poor}.

One of the major efforts in a \textit{simultaneous wireless information and power transfer} (SWIPT) architecture is to increase the \textit{direct-current} (DC) power at the output of the harvester without increasing transmit power. In the literature, this problem is approached by considering a constraint on the minimum delivery power at the output of the \textit{energy harvester} (EH). The EH, known as rectenna, is composed of an antenna followed by a rectifier.\footnote{In the literature, the rectifier is usually considered as a nonlinear device (usually a diode) followed by a low-pass filter. The diode is the main source of nonlinearity induced in the system.\label{footnote_1}} In \cite{Trotter_Griffin_Durgin_2009,Clerckx_Bayguzina_2016}, it is shown that the RF-to-DC conversion efficiency (harvested DC power over the power of received RF signal) is a function of rectenna's structure, as well as its input waveform (power and shape). Accordingly, in order to maximize rectenna's DC power output, a systematic waveform design is crucial to make the best use of an available RF spectrum \cite{Clerckx_Bayguzina_2016}. In \cite{Clerckx_Bayguzina_2016}, an analytical model for the rectenna's output is introduced via the Taylor expansion of the diode characteristic function and a systematic design for multisine waveform is derived. The nonlinear model and the design of the waveform was validated using circuit simulations in \cite{Clerckx_Bayguzina_2016, Clerckx_Bayguzina_2017} and recently confirmed through prototyping and experimentation in \cite{Kim_Clerckx_Mitcheson_arxive}. Those works also confirm the inaccuracy and inefficiency of a linear model of the rectifier\footnote{In the linear model, the output power of the EH, is proportional to the second moment of the received signal. The linear model has for consequence that the RF-to-DC conversion efficiency of the EH is constant and independent of the harvester's input waveform (power and shape) \cite{Zhang_Keong_2013,Zeng_Clerckx_Zhang2017}.\label{footnote_2}}.

The SWIPT literature has so far focused on the linear model of the rectifier, e.g., \cite{Grover_Sahai_2010,Zhang_Keong_2013,Park_Clerckx_2013}, whereas, considering nonlinearity effect changes the SWIPT design, signalling and architecture, significantly. In \cite{Clerckx_2016}, the SWIPT waveforms design and the characterization of achievable \textit{rate-power} (RP) region are studied on \textit{additive white Gaussian noise} (AWGN) channels accounting for the rectenna's nonlinearity with a power splitter at the receiver. In single-carrier transmission, it is shown that modulation with \textit{circular symmetric complex Gaussian} (CSCG) input is beneficial to delivered power compared to an unmodulated continuous wave. In multi-carrier transmission, however, it is shown that a non-zero mean Gaussian input distribution leads to an enlarged RP region compared to a CSCG input. This highlights that the choice of a suitable input distribution (and therefore modulation and waveform) for SWIPT is affected by the rectifier nonlinearity and motivates the study of the capacity of AWGN channels under nonlinear power constraints.

The capacity of complex and real discrete-time memoryless AWGN channels has been investigated in the literature under various constraints, extensively. The most classical one is the channel input average power constraint, under which the optimal input is demonstrated to be Gaussian distributed \cite{Shannon_1948}. It seems that the linear AWGN channel subject to transmit average power constraint is an exception and under many other constraints, the optimal input leads to discrete inputs. To mention a few, Smith in \cite{Smith:IC:71} considered a real AWGN channel with an average power and an amplitude constrained inputs, where he established that the optimal capacity achieving input distribution is discrete with a finite number of mass points. Similar results were reported in \cite{Shamai_BarDavid_1995} for complex AWGN channels with average and peak-power constraints and in \cite{AbouFaycal_Trott_Shamai_2001} for complex Rayleigh-fading channel under average power constraint with no \textit{channel state information} (CSI) at both the receiver and the transmitter. As a more general result, in \cite{Tchamkerten_2004}, a real channel is considered, in which sufficient conditions for the additive noise are provided, such that the support of the optimal bounded input has a finite number of mass points. In \cite{Fahs_AbouFaycal_2012}, real AWGN channels with nonlinear inputs are considered subject to multiple types of constraints such as the even moments and/or compact-support constraints, under which the optimal input is proved to be discrete with a finite number of mass points in the vast majority of the cases.

A survey of the literature reveals that almost all models considered for AWGN channels do not include the inevitable nonlinearities, such as fibre optic channels, power amplifiers or EHs. The lack of fundamental results in the literature relating to nonlinear models is becoming more sensible due to the growth of applications involving devices with nonlinear responses. The typical and straightforward approaches to tackle such problems are either considering linearized models with nonlinear effects being part of the noise model \cite[Sec 14.2]{desurvire_2009} or obtaining approximations and lower bounds on capacity by assuming Gaussian statistics \cite{Lomnitz_Feder, Mitra_Stark_2001}. As one of the novel works in the information theory literature, in \cite{Fahs_AbouFaycal_2012}, the authors consider a real AWGN channel with their focus on nonlinear channel inputs and different types of transmit power constraints.

Leveraging the aforementioned observations, we provide a step closer at identifying the fundamental limits of SWIPT structures taking into account the nonlinearities of the EH, i.e., rectenna. In this paper, we study a deterministic, complex and discrete-time memoryless AWGN channel under the transmit average power and amplitude constraints as well as a constraint on the delivered power at the output of the EH. We show that the delivered power, modelled as in \cite{Clerckx_Bayguzina_2016}, can be lower bounded by a sum of even moments of the baseband channel input. Motivated by this, we model the constraint on the delivered power as the (statistical) average over a linear combination of even moments of the channel input. The contributions of this paper are listed below.
\begin{itemize}
\item First, we show that the capacity of an AWGN channel under a transmit average power and a receiver delivered power constraints (this constraint is modelled as the average of linear combination of even moments of the channel input being larger than a threshold) is the same as the capacity of an AWGN channel. However, depending on the two constraints, the capacity can be either achieved using a unique CSCG input or approached arbitrarily (irrespective of the delivered power constraint) using time sharing between a Gaussian distribution and an on-off keying (OOK) distribution (with low probability of high amplitude signals).

\item Second, we show that under input average power and amplitude and output delivered power constraints, similarly to the results reported in \cite{Smith:IC:71,Shamai_BarDavid_1995,AbouFaycal_Trott_Shamai_2001} and \cite{Fahs_AbouFaycal_2012}, the capacity achieving input distribution is discrete in amplitude with a finite number of mass-points and with a uniformly distributed independent phase.

\item Third, as an application of the obtained results, we consider SWIPT over a complex AWGN channel, where the receiver is equipped with a rectenna in order to harvest and convert RF power into DC power. Taking the advantage of the small-signal approximation for rectenna's nonlinear output introduced in \cite{Clerckx_Bayguzina_2016,Clerckx_2016}, we obtain the general form of the delivered power for \textit{independent and identically distributed} (iid) complex inputs in terms of system baseband parameters. Assuming that the receiver jointly extracts information and harvests power from the received RF signal, it is shown that the delivered power at the receiver depends on different moments of the channel input. Defining RP region for the considered application, we obtain two inner bounds for the RP region. The first inner bound is based on merely iid complex Gaussian inputs, where we show that the optimal complex Gaussian inputs are zero mean. We also recognize a tradeoff between transmitted information and delivered power resulting from asymmetric power allocations between inphase and quadrature subchannels. The second inner bound is based on restricting the delivered power constraint and obtaining the necessary and sufficient conditions for optimality in the corresponding optimization probability space. Using numerical programming, it is observed that the Numerically Obtained Input distributions outperform their Gaussian counterparts.

\item Fourth, the analysis provides new engineering guidelines and refreshing views on the crucial role played by nonlinearity in SWIPT design. First, in contrast with the conventional linear model of the EH for which CSCG inputs are capacity achieving under average power constraints \cite{Varshney_2008,Grover_Sahai_2010}, CSCG inputs cannot achieve the optimal RP region boundaries in the presence of nonlinearity. Second, the EH nonlinearity enlarges the RP region. Hence, in contrast with other systems subject to nonlinear responses, where nonlinearity is compensated (e.g. \cite{Essiambre_Kramer_Winzer_Foschini_Goebel}), the nonlinearity in SWIPT is exploitable in the signal and system design and is beneficial to the system performance. Third, in contrast with the linear model for which time sharing between power and information transmission is suboptimal \cite{Zhang_Keong_2013}, time sharing between a CSCG distribution and an OOK signalling (with low probability of the On signal), is sufficient to approach the capacity in the presence of nonlinearity. Fourth, the efficacy of the derived and optimized input distributions to boost the harvested DC power is validated and confirmed through realistic circuit simulations. This sheds light on a new form of signal design for \textit{wireless power transfer} (WPT) relying on (energy) modulation for single-carrier transmission, as an alternative to the multi-carrier (energy) waveform approach of \cite{Clerckx_Bayguzina_2016}.

\item Fifth, as an independent result, we note that in analyzing complex AWGN channels, Bessel modified function of first kind of order zero appears frequently. Due to the form of Bessel functions, it is sometimes hard to analyse such channels. Accordingly, we obtain a tight upper bound on the Bessel modified function of first kind of order zero, which might also come useful in future applications and analysis.
\end{itemize}

\textit{Organization}: In Section \ref{Sec_System_Model}, we introduce the system model and define the channel capacity problem studied here. In Section \ref{Sec_Main_res}, we introduce the main results of the paper. A SWIPT problem is considered in Section \ref{Sec_Application} as an application of the main results introduced in Section \ref{Sec_Main_res}. In Section \ref{Sec_Del_Pow}, the delivered power for the considered SWIPT problem is obtained in terms of channel baseband parameters for iid channel inputs accounting for small-signal approximations of rectenna. Defining the RP region in Section \ref{Sec_RP_Region}, an inner bound on the RP region based on complex Gaussian distributed inputs and an inner bound on the RP region based on the results developed in Section \ref{Sec_Main_res} are introduced in Section \ref{Sec_LB1} and Section \ref{Sec_LB2}, respectively. In Section \ref{Sec_Num_res}, numerically obtained inner bounds of RP region are illustrated. In Section \ref{sec_Disc}, some problems are posed as potential future research directions. We conclude the paper in Section \ref{Sec_Conclusion} and the proofs for some of the results are provided in the Appendices at the end of the paper.

\textit{Notations}: Throughout this paper, the standard CSCG distribution is denoted by $\mathcal{CN}(0,1)$. Complex conjugate of a complex number $c$ is denoted by $\overline{c}$. For a random process $X(t)$, the corresponding random variable at time index $k$ is represented by $\pmb{x}_k$. The support of the random variable $\pmb{x}_k$ is denoted as $\text{supp}\{\pmb{x}_k\}$. $\pmb{x}_r$ and $\pmb{x}_i$ denote the real and imaginary parts of the complex random variable $\pmb{x}$, respectively. $\text{Re}\{\cdot\}$ and $\text{Im}\{\cdot\}$ are real and imaginary operators, respectively. Define
\begin{align}\label{S_l}
a_l\triangleq\text{sinc}(l+1/2),
\end{align}
for integer $l$, where $\mathrm{sinc}(t)=\frac{\sin(\pi t)}{\pi t}$. $F_{\pmb{x}}(x)$ and $f_{\pmb{x}}(x)$ denote, respectively, the cumulative distribution function (cdf) and the probability density function (pdf) of the random variable $\pmb{x}$. For the random process $X(t)$, the expectation over statistical randomness $\mathbb{E}[\cdot]$ and averaging over time $\mathcal{E}[\cdot]$ are defined as
\begin{align}\label{E91}
\mathbb{E}[X(t)]&=\int_{-\infty}^{\infty}x(t)dF_{\pmb{X}(t)}(x),\\
\mathcal{E}[X(t)]&=\lim\limits_{T\rightarrow\infty}\frac{1}{2T}\int_{-T}^{T}X(t) dt,
\end{align}
respectively. $\Phi(\cdot,\cdot;\cdot)$ denotes the confluent hypergeometric function defined as in \cite[Section 9.21]{gradshteyn2007}. The Heaviside step function is denoted by $U(x)$, and the error function is defined as $\mathrm{erf}(x)=2/\sqrt{\pi}\int_{0}^{x}e^{-t^2}dt$.

\section{System Model, Problem Definition and Preliminaries}\label{Sec_System_Model}
Consider the following complex representation of a discrete-time memoryless AWGN channel,
\begin{align}\label{E69}
\pmb{y}_k=\pmb{x}_k+\pmb{n}_k,
\end{align}
where $\{\pmb{y}_k\}$, $\{\pmb{x}_k\}$ and $\{\pmb{n}_k\}$ represent the sequences of complex-valued samples of the channel output, input and AWGN, respectively, and $k$ is the discrete-time index. The real and imaginary parts of the signal $\{\pmb{y}_k\}$ indicate the inphase and quadrature components, respectively. The noise samples $\{\pmb{n}_k\}$ are assumed to be CSCG distributed as $\mathcal{CN}(0,2)$, i.e., $\mathbb{E}[\text{Re}\{\pmb{n}_k\}^2]=\mathbb{E}[\text{Im}\{\pmb{n}_k\}^2]=1$ and $\mathbb{E}[\text{Re}\{\pmb{n}_k\}\text{Im}\{\pmb{n}_k\}]=0$\footnote{We choose $\mathbb{E}[|\pmb{n}_k|^2]=2$ for brevity, however the results can be extended to any noise variance value.\label{footnote_3}}.

We are interested in the capacity of the channel in (\ref{E69}) with input samples subject to
\begin{align}\label{E1}
\left\{\begin{array}{lll}
&\mathbb{E}[|\pmb{x}_k|^2]  \leq  P_a\\
&P_{d} \leq \mathbb{E}[g(|\pmb{x}_k|)]   \\
&|\pmb{x}_k|\leq r_p
\end{array}\right.,
\end{align}
for all $k$, where throughout the paper $P_a<\infty$, $P_d<\infty$ and $r_p\leq\infty$ are interpreted as the transmitter maximum allowable average power, minimum delivered power and channel input amplitude constraints, respectively.
\begin{rem}
Note that the single-letter characterizations of the constraints in (\ref{E1}) are equivalent to their multi-letter characterization in the operational definition of capacity. The details have been provided in Lemma \ref{Lem_single} Appendix \ref{Guide}.
\end{rem}
Throughout the paper, any operator that involves a random variable reads with the term \textit{almost-surely} (e.g. $|\pmb{x}_k|\overset{\text{a.s.}}\leq r_p$).  $g(\cdot)$ is assumed to be a continuous positive function having the form of
\begin{align}\label{E54}
g(r)=\sum\limits_{i=0}^{d}\alpha_i r^{2i},~r\geq 0,
\end{align}
where $d\geq 2$ is an arbitrary integer. Note that since $g(r)$ is assumed to be a positive function, we have $\alpha_d>0$, and hence, $\lim_{r\rightarrow\infty}g(r)=\infty$\footnote{In Section \ref{Sec_Application}, we show that the delivered power based on the experimentally validated model in \cite{Clerckx_Bayguzina_2016} can be lower bounded by even moments of the baseband channel input. Motivated by this, we model the delivered power constraint as the average over function (\ref{E54}).\label{footnote_4}}.

\begin{rem}\label{Rem0}
The scenario $g(r)=\alpha_0+\alpha_1r^2$ is not considered in this paper, as the capacity problem in (\ref{E1}) boils down to either \cite{Shannon_1948} (when $r_p=\infty$), where a CSCG distribution is optimal, or \cite{Shamai_BarDavid_1995} (when $r_p<\infty$), where optimal distribution is discrete with a finite number of mass points. Accordingly, we are interested in $g(r)$ with $\alpha_i\neq0$ for at least one of $i=2,\ldots,n$.
\end{rem}

The capacity of a discrete-time memoryless complex AWGN channel \cite[Chapter 7]{Gallager_book} is therefore given by\footnote{Given that the channel model is stationary and memoryless, and due to the type of the constraints, it can be verified that the capacity-achieving statistics of $\pmb{x}_k$ are also memoryless; therefore, we can suppress the time index.\label{footnote_5}}
\begin{equation}\label{E3}
\begin{aligned}
C(P_a,P_d,r_p)=& \sup\limits_{f_{\pmb{x}}(x)}
& & I(\pmb{x};\pmb{y}) \\
& \text{s.t.}
& & \left\{\begin{array}{lll}
\mathbb{E}[|\pmb{x}|^2]  \leq  P_a, \\
P_d\leq \mathbb{E}[g(|\pmb{x}|)]  ,\\
|\pmb{x}|\leq r_p,
\end{array}\right.
\end{aligned}
\end{equation}
By expressing $I(\pmb{x};\pmb{y})$ in terms of differential entropies, i.e., $I(\pmb{x};\pmb{y})=h(\pmb{y})-\ln 2\pi e$, (\ref{E3}) boils down to the supremization of differential entropy $h(\pmb{y})$. Using the polar coordinates\footnote{The polar representation simplifies the problem, since the constraints are circular symmetric.\label{footnote_6}} $\pmb{x}=\pmb{r}e^{i\pmb{\theta}}$ and $\pmb{y}=\pmb{R}e^{i\pmb{\phi}}$ ($\pmb{r},\pmb{R}\geq 0$ and $\pmb{\theta},\pmb{\phi}\in[-\pi,\pi)$) and following the same steps in \cite[eq. 5 to eq. 12]{Shamai_BarDavid_1995}, we have
\begin{align}\label{E4}
h(\pmb{y})\leq -\int\limits_{0}^{\infty} f_{\pmb{R}}(R;F_{\pmb{r}})\ln \frac{f_{\pmb{R}}(R;F_{\pmb{r}})}{R} dR +\ln 2\pi,
\end{align}
where $f_{\pmb{R}}(R;F_{\pmb{r}})$ is the pdf of $\pmb{R}$ induced by $F_{\pmb{r}}$ and is given by
\begin{align}\label{E93}
f_{\pmb{R}}(R;F_{\pmb{r}})=\int\limits_{0}^{r_p}K(R,r)dF_{\pmb{r}}(r),
\end{align}
where the kernel $K(R,r)$ is defined as
\begin{align}\label{E92}
K(R,r)\triangleq R e^{-\frac{R^2+r^2}{2}} I_0(rR),
\end{align}
with $I_0(x)=1/\pi\int_{0}^{\pi}e^{x\cos(\theta)}d\theta$ the modified Bessel function of the first kind and order zero. Note that by selecting $\pmb{r}$ and $\pmb{\theta}$ independent with uniformly distributed $\pmb{\theta}$ over $[-\pi,\pi)$\footnote{Note that this causes no loss of optimality, since the constraints are circular symmetric.\label{footnote_7}}, (\ref{E4}) holds with equality and we have
\begin{align}\label{Mr1}
f_{\pmb{R},\pmb{\phi}}(R,\phi)=\frac{1}{2\pi}f_{\pmb{R}}(R;F_{\pmb{r}}).
\end{align}
Therefore, the optimization problem in (\ref{E3}) is reduced to the following problem
\begin{equation}
\begin{aligned}\label{E19}
C(P_a,P_d,r_p)=& \underset{F_{\pmb{r}}\in \Omega_1\cap\Omega_2}\sup
& & H(F_{\pmb{r}})-\ln e ,
\end{aligned}
\end{equation}
where $F_{\pmb{r}}(0^{-})=0$, $F_{\pmb{r}}(r_p)=1$ and $H(F_{\pmb{r}})$, $\Omega_1$ and $\Omega_2$ are given as
\begin{align}\label{E8}
  H(F_{\pmb{r}})&\triangleq -\int\limits_{0}^{\infty} f_{\pmb{R}}(R;F_{\pmb{r}})\ln \frac{f_{\pmb{R}}(R;F_{\pmb{r}})}{R} dR,
\end{align}
and
\begin{subequations}\label{E47}
\begin{align}\label{E47a}
  \Omega_1&=\left\{F_{\pmb{r}}:\int\limits_{0}^{r_p}r^2dF_{\pmb{r}}(r)\leq P_a\right\},\\\label{E47b}
  \Omega_2&=\left\{F_{\pmb{r}}:P_d\leq \int\limits_{0}^{r_p}g(r)dF_{\pmb{r}}(r) \right\}.
\end{align}
\end{subequations}

\section{Main results}\label{Sec_Main_res}
In this section, we provide the main results of this paper. First, we characterize the capacity in (\ref{E19}) when the channel input amplitude constraint is $r_p=\infty$. In the following theorem, we study the capacity problem in (\ref{E19}), when $r_p<\infty$. We accordingly, derive the necessary and sufficient condition for the optimal distributions achieving the capacity.

\begin{thm}\label{T0}
The capacity of the channel in (\ref{E69}) for $r_p=\infty$ is
\begin{align}\label{E7}
 C(P_a,P_d,\infty) =\ln\left(1+\frac{P_a}{2}\right).
\end{align}
Let $P_G=1/P_a\int_{0}^{\infty}rg(r)e^{-\frac{r^2}{2P_a}}dr$ be the delivered power corresponding to an input distributed as $\pmb{x}\sim\mathcal{CN}(0,P_a)$. The supremum in (\ref{E3}) is achieved\footnote{i.e., $\sup$ can be replaced by $\max$. \label{footnote_8}} by a unique input if and only if $P_d \leq P_G$, in which case it is $\pmb{x}\sim\mathcal{CN}(0,P_a)$. If $P_d> P_G$, (\ref{E7}) is not attained, however, (\ref{E7}) can be approached arbitrarily closely by using time-sharing between a Gaussian distribution and OOK.
\end{thm}
\textit{Proof}: See Appendix \ref{A0}.

\begin{rem}
One example of a sequence of distributions (approaching (\ref{E7}) arbitrarily closely) is illustrated in (\ref{E80}), which is a time sharing between a CSCG and an OOK (with a low probability for the On signal) distributed inputs.
\end{rem}

\begin{figure}
\begin{centering}
\includegraphics[scale=0.25]{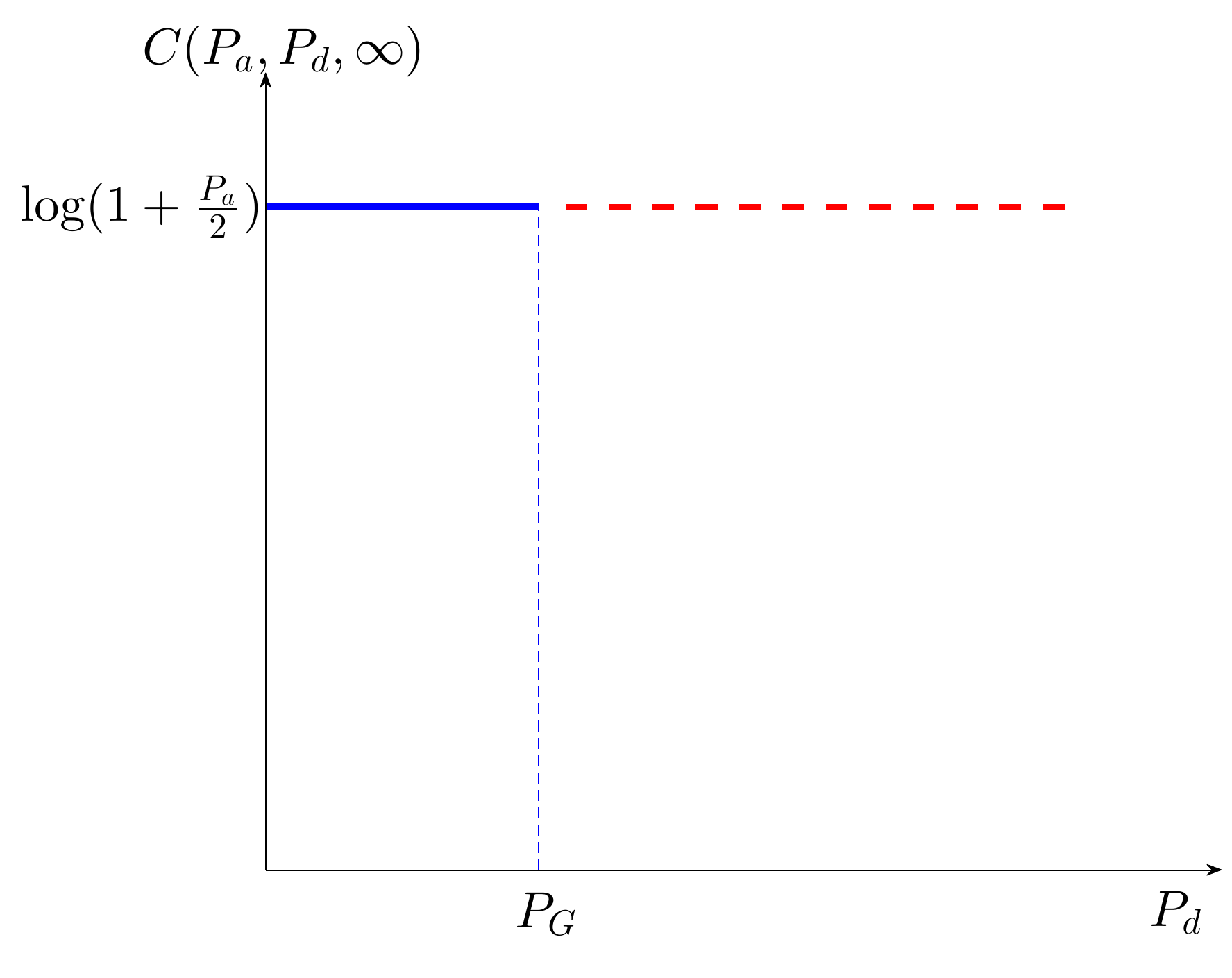}
\caption{The capacity $C(P_a,P_d,\infty)$ of an AWGN channel. The solid blue line is achievable by a unique input $\pmb{x}\sim\mathcal{CN}(0,P_a)$, however, the red dashed line can be approached.}\label{F0}
\par\end{centering}
\vspace{0mm}
\end{figure}

From Theorem \ref{T0}, it is verified that for $n\geq 2$ in (\ref{E54}), the capacity of an AWGN channel in (\ref{E69}) for $r_p=\infty$ is independent of the value of the delivered power constraint, i.e., $P_d$. That is, given $P_a$, the capacity $C(P_a,P_d,\infty)$ is constant with $P_d$. This is represented in Figure \ref{F0}, where the solid line illustrates the capacity $C(P_a,P_d,\infty)$ achievable by $\pmb{x}\sim\mathcal{CN}(0,P_a)$, and the dashed line illustrates the capacity $C(P_a,P_d,\infty)$ that can be approached arbitrarily using time sharing\footnote{We note that power splitting at the receiver (dividing the received signal into two streams with different power levels using a power splitter) results in a larger RP region \cite{Zhang_Keong_2013}. The main use of time sharing in our results is that from approaching the capacity point of view, time sharing is sufficient. We also note that, in a practical receiver, the information decoder and the EH are separate. Accordingly, in practical applications, power splitting may still be preferred.\label{footnote_9}} between high information distributions, e.g. CSCG inputs, and high power distributions, e.g. OOK inputs, (see Appendix \ref{A0} for construction of such inputs).

Note that, the result of Theorem \ref{T0} is due to the fact that the function $g(r)$ is of the order of at least $4$. In Section \ref{Sec_Application}, we show that accounting for the rectifier nonlinearity\footnote{We note that, in practice, nonlinearity of the EH (rectenna) occurs in the low average RF input power regime.\label{footnote_10}} at the receiver, the delivered power at the output of the EH depends on higher order moments of the channel input $\pmb{x}$.

\begin{thm}\label{T1}
The optimal distribution denoted by $F_{\pmb{r}^{o}}$ achieving the capacity $C(P_a,P_d,r_p)$ for $r_p<\infty$, is unique and its corresponding set of points of increase\footnote{$x$ is said to be a point of increase of $F_{\pmb{x}}$ if and only if $\textrm{Pr}(x-\eta<\pmb{x}<x+\eta)>0$ for all $\eta>0$.\label{footnote_11}} is finite (the cardinality of the support of the random variable $\pmb{r}^{o}$ is finite, i.e., $|\text{supp}\{\pmb{r}^{o}\}|<\infty$). Furthermore, $F_{\pmb{r}^{o}}$ is optimal if and only if there exist unique parameters $\lambda\geq 0$ and $\mu\geq 0$ for which
\begin{subequations}\label{E70}
\begin{align}
&h(r;F_{\pmb{r}^{o}})-\lambda r^2\!+\!\mu g(r) \!-\!K_{F_{\pmb{r}^o}}=0,~\forall r\in   \emph{supp}\{\pmb{r}^{o}\},\\
&h(r;F_{\pmb{r}^{o}})-\lambda r^2\!+\!\mu g(r) \!-\!K_{F_{\pmb{r}^o}}\leq 0,~\forall r\in [0,r_p],
\end{align}
\end{subequations}
where $K_{F_{\pmb{r}^o}}\triangleq H(F_{\pmb{r}^o})-\lambda P_a+\mu P_d$ and
\begin{align}\label{E14}
h(r;F_{\pmb{r}^o})&=-\int\limits_{0}^{\infty} K(R,r) \ln{\frac{f_{\pmb{R}}(R;F_{\pmb{r}^o})}{R}}dR.
\end{align}
\end{thm}
\textit{Proof}: See Appendix \ref{A1}.

Note that the results in (\ref{E70}) are important in the sense that they can be utilized to obtain the optimal distributions using numerical programming. In \cite{Morsi_Jamali}, the capacity of a real AWGN channel is studied with $g(r)=I_0(r)$. It can be easily verified that for both real and complex AWGN channels the obtained results (uniqueness and finite cardinality of the optimal input distribution) in \cite{Morsi_Jamali} and here in Theorem \ref{T1} remain valid if the function $g(r)$ grows faster than $r^2$, i.e., $r^2=\mathcal{O}(g(r))$\footnote{By definition, given two functions $f(\cdot)$ and $g(\cdot)$, we write $f(x)=\mathcal{O}(g(x))$ if and only if there exist two positive scalars, $c>0,~x_0>0$, such that $|f(x)|\leq c|g(x)| , \forall x>x_0$.\label{footnote_12}}.

\begin{rem}\label{Rem1}
Rewriting the KKT condition for the inequality in (\ref{E70}), we get
\begin{align}\label{E2}
0\leq \mu\leq \frac{K_{F_{\pmb{r}^o}}+2+\lambda r^2}{g(r)},~r\in [0,r_p],
\end{align}
where we used the inequality $h(r;F_{\pmb{r}})\geq -2$ for any $F_{\pmb{r}}\in \Omega_1\cap \Omega_2$ (see (\ref{E59}) in Appendix \ref{A1}). We note that, since by definition the function $g(r)$ grows faster than $r^2$, we have $\mu\rightarrow 0$ as $r_p\rightarrow \infty$. The intuition behind this is as follows. $\mu$ can be considered as the opposite sign of $\partial C(P_a,P_d,r_p)/\partial r_p$. As $r_p$ increases, $C(P_a,P_d,r_p)$ approaches $C(P_a,P_d,\infty)$. From Theorem \ref{T0}, we already know that capacity $C(P_a,P_d,\infty)$ is unchanged for any $P_d< \infty$. Therefore, $\partial C(P_a,P_d,r_p)/\partial r_p\rightarrow0$, and accordingly,  $\mu\rightarrow0 $ as $r_p$ increases. In other words, the dependency of the capacity on $r_p$ reduces as $r_p$ grows large.
\end{rem}

\begin{rem}
In \cite[Corollary 2]{Morsi_Jamali}, it is stated that for a real AWGN channel and $g(r)=I_0(r)$, when $r_p\rightarrow \infty$ and $P_d$ is greater than the feasible delivered power corresponding to Gaussian input, the capacity is still achievable and the corresponding input distribution is discrete with a finite number of mass points. We note that, this claim cannot hold, since as in Theorem \ref{T0}, the capacity is not achievable, however, it can be approached arbitrarily (See Appendix \ref{A0}, for construction of such distributions approaching capacity when $r_p=\infty$.).
\end{rem}


\section{Application}\label{Sec_Application}
As an application of the results in Section \ref{Sec_Main_res}, in this section, we consider the channel in (\ref{E69}), under a scenario where the receiver is equipped with a nonlinear EH. In the following, we first explain the transmission process. Next, we obtain a baseband equivalent for the harvested power at the receiver. Later, we define the rate-power region, and obtain two inner bounds on the rate-power region.

\textit{Transmitter}:
The transmitted process $X(t)$ is produced as
\begin{align}\label{E90}
X(t)=\sum_{k}\pmb{x}_k\text{sinc}(f_wt-k),
\end{align}
where $\pmb{x}_k$ is an information-power symbol at time index $k$, modelled as a random variable, which is produced in an iid fashion\footnote{In this paper, we are interested in a random coding achievability bound in which iid symbols are mapped into continuous waveform through sync interpolation. Accordingly, the iid assumption on transmitted symbols $\pmb{x}_k$ in (\ref{E90}) is an imposed condition; and it is due to the fact that in general, deriving the baseband equivalent of the delivered power (see equation (\ref{eqn:1})) is cumbersome.\label{footnote_13}}. Next, the process $X(t)$ is upconverted to the carrier frequency $f_c$ and is sent over the channel.

\textit{Receiver}:
The filtered received RF waveform at the receiver is modelled as
\begin{align}\label{E96}
  Y_{\text{rf}}(t) &=\sqrt{2}\text{Re}\left\{Y(t)e^{j2\pi f_ct}\right\},
\end{align}
where $Y(t)$ is bandlimited to $f_w/2$ Hz. In order to have a narrowband transmission, we assume that $f_c\gg2f_w$.

\textit{Power}: At the receiver, the power of the RF signal $Y_{\text{rf}}(t) $ is captured via the rectenna. Leveraging the small-signal approximation for rectenna's output introduced in \cite{Clerckx_Bayguzina_2016,Clerckx_2016},\footnote{According to \cite{Clerckx_Bayguzina_2016}, due to the presence of a diode in rectenna's structure, its output current is an exponential function, which is approximated by expanding its Taylor series. The approximation used here, is the fourth moment truncation of Taylor series, in which the first and third moments are zero with respect to the time averaging. Discussions on the assumptions and validity of this model can be found in \cite{Clerckx_Bayguzina_2016}.\label{footnote_14}} the delivered power, denoted by $P_{\text{del}}$ is modelled as\footnote{According to \cite{Clerckx_Bayguzina_2016}, rectenna's output in (\ref{eqn:1}) is in the form of current with unit Ampere. However, since power is proportional to current, with abuse of notation, we refer to the term in (\ref{eqn:1}) as power.\label{footnote_15}}
\begin{align}\label{eqn:1}
P_{\text{del}}=\mathbb{E}\mathcal{E}[k_2Y_{\text{rf}}(t)^2 + k_4 Y_{\text{rf}}(t)^4],
\end{align}
where $k_2$ and $k_4$ are constants. Note that, in the linear model for the delivered power $P_{\text{del}}$, in (\ref{eqn:1}), we have only the second moment of the received RF signal $Y_{\text{rf}}(t)$, where the optimal input is shown to be a CSCG distribution \cite{Shannon_1948}.

\textit{Information}: The signal $Y_{\text{rf}}(t) $ is downconverted producing the baseband signal $Y(t)$ given as\footnote{We model the baseband equivalent channel impulse response as $H(\tau,t)=\sum_{i}\delta(\tau)+W(t)$, where the delay and the gain of the channel are assumed to be $0$ and $1$, respectively.\label{footnote_16}}
\begin{align}
Y(t)=X(t)+W(t).
\end{align}
Next, $Y(t)$ is sampled with a sampling frequency $f_w$ producing $\pmb{y}=\pmb{x}+\pmb{n}$ as in (\ref{E69}).\footnote{Due to the assumption of iid channel inputs and discrete memoryless channel, we neglect the time index $k$.\label{footnote_17}}

\subsection{Delivered power in the baseband}\label{Sec_Del_Pow}
From a communications system design point of view, it is most preferable to have baseband equivalent representation of the system. Henceforth, in the following Proposition, we derive the delivered power $P_{\text{del}}$ at the receiver (see (\ref{eqn:1})) in terms of the system baseband parameters.

\begin{lem}\label{Prop1}
Assuming the channel input distributions are iid, the delivered power $P_{\text{del}}$ at the receiver can be expressed as
\begin{align}\label{eqn:22}
P_{\text{del}}=\alpha (Q+\tilde{Q})+\beta P+\gamma,
\end{align}
where $\tilde{Q}$ is given by
\begin{align}\nonumber
\tilde{Q}&=\frac{1}{3}\big(Q_{r}+Q_{i}+2(\mu_{r}T_{r}+\mu_{i}T_{i})\\
&+6P_{r}P_{i}+6P_{r}(P_{r}-\mu_{r}^{2})+6P_{i}(P_{i}-\mu_{i}^{2})\big),
\end{align}
and the parameters $\alpha,~\beta$ and $\gamma$ are given as
\begin{align}\label{eqn:32}
\alpha&=\frac{3k_4}{2},\\
 \beta&=k_2+48k_4, \\
 \gamma&=4k_2+96k_4,
\end{align}
where $k_2,~k_4$ are constants as in (\ref{eqn:1}), and $Q=\mathbb{E}[|\pmb{x}|^4]$, $T=\mathbb{E}[|\pmb{x}|^3]$, $P=\mathbb{E}[|\pmb{x}|^2]$, $\mu=\mathbb{E}[\pmb{x}]$. Similarly, $Q_r=\mathbb{E}[\pmb{x}_r^4]$, $T_r=\mathbb{E}[\pmb{x}_r^3]$, $P_r=\mathbb{E}[\pmb{x}_r^2]$, $\mu_r=\mathbb{E}[\pmb{x}_r]$ and $Q_i=\mathbb{E}[\pmb{x}_i^4]$, $T_i=\mathbb{E}[\pmb{x}_i^3]$, $P_i=\mathbb{E}[\pmb{x}_i^2]$, $\mu_i=\mathbb{E}[\pmb{x}_i]$.
\end{lem}
\textit{Proof}: See Appendix \ref{app:1}.

It is observed that the obtained delivered power $P_{\text{del}}$ in (\ref{eqn:22}) is a function of different (odd and even) moments of the channel input. In the next section, we show that a convex (with respect to the input distribution) lower bound of the delivered power $P_{\text{del}}$ is obtained by restricting $P_{\text{del}}$ to merely the even moments of the channel input. Accordingly, we use the restricted version of the delivered power to use the results in Theorem \ref{T0}.

\begin{rem}
We note that, obtaining a closed form expression for the delivered power $P_{\text{del}}$ at the receiver, when the channel inputs are not iid is cumbersome. This is due to the fact that the fourth moment of the received RF signal $Y_{\text{rf}}(t)$ creates dependencies of the statistics of the present channel input on the statistics of the channel inputs on the other time indices (see e.g., eq. (\ref{eqn:9}) and eq. (\ref{eqn:30}) in Appendix \ref{app:1}).
\end{rem}

\subsection{Rate-Power (RP) region}\label{Sec_RP_Region}
We define the RP region as the convex hull of the following union of regions
\begin{align}\label{E71}
\mathcal{R}(P_a,r_p)\!=\!\bigcup\limits_{P_d} \left\{(R,P)\!:\! R<C_{\text{SWIPT}}(P_a,P_d,r_p),P\leq P_d\right\},
\end{align}
where $C_{\text{SWIPT}}(P_a,P_d,r_p)$ is defined similarly to (\ref{E3}) as
\begin{equation}\label{E72}
\begin{aligned}
C_{\text{SWIPT}}(P_a,P_d,r_p)=& \sup
& & I(\pmb{x};\pmb{y}) \\
& f_{\pmb{x}}(x):
& & \left\{\begin{array}{lll}
\mathbb{E}[|\pmb{x}|^2]  \leq  P_a, \\
P_d\leq P_{\text{del}} ,\\
|\pmb{x}|\leq r_p,
\end{array}\right.
\end{aligned}
\end{equation}
and $P_{\text{del}} $ is given in (\ref{eqn:22}).


In the following, we consider two different inner bounds on the RP region defined in (\ref{E71}). In the first approach, we assume that the inputs are Gaussian distributed, where it is shown that the optimal Gaussian inputs are zero mean. In the second, we obtain an inner bound on the harvested power in (\ref{eqn:22}) by considering a convex subset of optimization probability space, and accordingly, apply the result of Theorem \ref{T1}.

\subsubsection{Complex Gaussian Inputs}\label{Sec_LB1}

Assuming that the inputs are Gaussian distributed, we show that for the considered scenario, there is a tradeoff between the rate of the transmitted information, namely $I(\pmb{x};\pmb{y})$ and delivered power $P_{\text{del}}$ at the receiver, and accordingly, we characterize the tradeoff.

\begin{lem}\label{Prop2}
If a channel input distribution $f_{\pmb{x}}(x)$ is complex Gaussian, the supremum in (\ref{E72}) is achieved by zero mean inputs, i.e., $\text{Re}\{\pmb{x}\}\sim\mathcal{N} (0,P_r)$, and $\text{Im}\{\pmb{x}\}\sim \mathcal{N}(0,P_i)$, where $P_r+P_i=P_a$. Furthermore, let $P_{\text{del,max}}=3\alpha {P_{a}}^2+2\beta P_{a} +\gamma$ and $P_{\text{del,min}}=2\alpha {P_{a}}^2+2\beta P_{a} +\gamma$ be the maximum and minimum delivered power at the receiver, respectively. If $P_{d}>P_{\text{del,max}}$, the solution does not exist. If $P_{d}=P_{\text{del,max}}$, the maximum in (\ref{E72}) is attained by $P_i=0,P_r=P_{a}$ or $P_i=P_a,P_r=0$. If $P_{\text{del,min}}<P_{d}< P_{\text{del,max}}$, the optimal power allocation that attains the maximum in (\ref{E72}) is given by $P_i^*$ and $P_r^*=P_a-P_i^*$, where $P_i^*$ is chosen, such that the following equation is satisfied
\begin{align}\label{eqn:31}
2\alpha (4P_i^{*2}+3P_a^2-8P_aP_i^{*})+2\beta P_a+\gamma=P_{d}.
\end{align}

For $P_{d}\leq P_{\text{del,min}}$, the optimal power allocation is attained by $P_i^*=P_r^*=P_{a}/2$ and the delivered power is still $P_{\text{del,min}}$.

\end{lem}
\textit{Proof}: See Appendix \ref{app:2}.

We note that the tradeoff between transmitted information and delivered power for Gaussian inputs, results from the asymmetric power allocation between inphase and quadrature subchannels. We have illustrated the RP region corresponding to Gaussian inputs in Section \ref{Sec_Num_res}.

\begin{rem}
From (\ref{eqn:22}), it is seen that the delivered power $P_{\text{del}}$ at the receiver depends on the average power $P_r,P_i$, as well as the fourth moment $Q_r,Q_i$ of the channel input $\pmb{x}$. This is due to the presence of the fourth moment of the received RF signal in modelling the rectenna's output. From Lemma \ref{Prop2}, it is seen that the maximum rate corresponding to $P_d=P_{\text{del,max}}$ is when the available power at the transmitter is fully allocated to one of the real or imaginary dimensions. This is because allocating power to one dimension, leads to a higher fourth moment statistic. On the other hand, the maximum rate corresponding to $P_d=P_{\text{del,min}}$ is when the available power is equally distributed between the real and the imaginary dimensions. Note that as also mentioned in Remark \ref{Rem0}, there is no tradeoff when the linear model is considered for the delivered power ,i.e., $n<2$ in (\ref{E54}).
\end{rem}

\subsubsection{Restricted optimization probability space}\label{Sec_LB2}
In this section, we consider an inner bound on the RP region defined in (\ref{E71}), by considering a convex subset of the optimization probability space in (\ref{E72}). The reason for restricting the optimization probability space is to utilize the analytic results presented in Section \ref{Sec_Main_res}. Note that the optimization probability space in (\ref{E72}) is convex with respect to the peak amplitude and average power constraint. Accordingly, we only need to consider a convex subset for the delivered power constraint. Then, the convexity of the optimization probability space follows due to the fact that the intersection of a finite number of convex sets is convex.

The convex subset of the optimization probability space in (\ref{E72}) is obtained by noting that the delivered power at the receiver in (\ref{eqn:1}) can be lower bounded by the sum of the even moment terms of its baseband equivalence in (\ref{eqn:22}) as below
\begin{align}\label{E75}
P_{\text{del}}&=k_2\mathbb{E}\left[|\pmb{y}_k|^2\right]+\frac{3k_4}{2}\left(\mathbb{E}[|\pmb{s}_{2k+1}|^2]+\mathbb{E}[|\pmb{s}_{2k}|^2]\right)\\
&>k_2\mathbb{E}\left[|\pmb{y}_k|^2\right]+\frac{3k_4}{2}\mathbb{E}[|\pmb{s}_{2k}|^2]\\\label{E82}
&=k_2\mathbb{E}\left[|\pmb{y}_k|^2\right]+\frac{3k_4}{2}\mathbb{E}[|\pmb{y}_{k}|^4]\\\nonumber
&=\frac{3k_4}{2}\mathbb{E}\left[|\pmb{x}_k|^4\right]+(k_2+24k_4)\mathbb{E}\left[|\pmb{x}_k|^2\right]\\\label{E83}
&\quad+4k_2+48k_4\\\label{E84}
&=\mathbb{E}[g_{\text{NL}}(\pmb{r})],
\end{align}
where in (\ref{E75}) we use the definition $\pmb{s}_k \triangleq |Y(k/2f_w)|^2$ and it is due to (\ref{E74}) (see Appendix \ref{app:1} for more details). (\ref{E82}) is due to (\ref{E85}). In (\ref{E84}), we have $\pmb{r}=|\pmb{x}|$ and $g_{\text{NL}}(r)$ is given as
\begin{align}\label{E81}
g_{\text{NL}}(r)=\frac{3k_4}{2}r^4+(k_2+24k_4)r^2+4k_2+48k_4.
\end{align}
By $g_{\text{NL}}(r)$ in hand and noting that $I(\pmb{x};\pmb{y})= H(F_{\pmb{r}})-1$, (\ref{E72}) can be written as
\begin{equation}\label{E76}
\begin{aligned}
C_{\text{IB}}(P_a,P_d,r_p)=& \sup
& &  H(F_{\pmb{r}})-1 \\
& F_{\pmb{r}}:
& & \left\{\begin{array}{lll}
\mathbb{E}[\pmb{r}^2]  \leq  P_a, \\
P_{d}\leq \mathbb{E}[g_{\text{NL}}(\pmb{r})]  ,\\
\pmb{r}\leq r_p.
\end{array}\right.
\end{aligned}
\end{equation}
The inner bound for the RP region in (\ref{E71}) is obtained by finding the corresponding delivered power $\mathbb{E}[g_{\text{NL}}(\pmb{r})]$ and transmitted information $I(\pmb{x};\pmb{y})$ of the optimal solutions of the problem (\ref{E76}). We illustrate the related results in Section \ref{Sec_Num_res}.

\section{Numerical Results}\label{Sec_Num_res}
In this section, we first illustrate through numerical evaluations the RP regions and highlight the benefits of nonlinear energy harvesting. We then evaluate through realistic circuit simulations the impact of various input distributions on the harvested DC power in WPT and contrast with the analytical results.

\subsection{Numerical Evaluations of SWIPT RP Regions}
In this section, we provide some numerical illustrations of the two inner bounds (see Section \ref{Sec_LB1} and \ref{Sec_LB2}) for the RP region defined in Section \ref{Sec_RP_Region}. In the following, we first summarize the steps in obtaining the bounds, and next, we illustrate the obtained numerical results.

\textit{Complex Gaussian inputs}: To obtain the RP region corresponding to Gaussian inputs, we use (\ref{eqn:31}). Note that when symmetric power allocation is used between the real and imaginary subchannels, i.e., $\mathbb{E}[\pmb{x}_i^2]=\mathbb{E}[\pmb{x}_r^2]=P_a/2$, the delivered power is $P_{\text{del},\min}$ with the transmitted information $\ln(1+P_a/2)$. We gradually increase $P_d$ ($P_d\geq P_{\text{del,min}}$) and using the fact that the average power constraint is satisfied with equality (see Lemma \ref{Prop1}) and using (\ref{eqn:31}), the optimal power allocations for inphase and quadrature channels are obtained. We continue increasing $P_d$ until allocated power for one of the subchannels gets zero. At this point, the delivered power is equal to $P_{\text{del,max}}$ and the transmitted information is $1/2\ln(1+P_a)$.

\textit{Inputs obtained by restricting the optimization probability space}: To obtain the RP region corresponding to the distributions obtained by solving (\ref{E76}), we resort to numerical programming. Accordingly, we solve the optimization problem in (\ref{E76}) using the interior-point algorithm implemented by the $\mathrm{fmincon}$ function in MATLAB software. Note that, since we already know that the optimal distribution is discrete with a finite number of mass points, the numerical optimization is over the position, the probabilities and the number of the mass points. Hence, there are $2m$ parameters to be optimized, where $m$ is the number of the mass points. We aim at calculating the capacity $I(\pmb{x};\pmb{y})$ in (\ref{E76}) under given an average power and an amplitude constraints and for different values of the delivered power constraint. As a result, we consider the following unconstraint optimization problem
\begin{align}\label{E78}
H(F_{\pmb{r}})-\lambda \mathbb{E}[\pmb{r}^2]+\mu \mathbb{E}[g_{\text{NL}}\left(\pmb{r}\right)],~0\leq \pmb{r}\leq r_p,~\lambda,\mu\geq 0.
\end{align}
In the following, the different steps of the optimization are summarized:
\begin{enumerate}
  \item Fix the average power constraint. Set $P_d =P_{\text{del,min}}+\delta$, where $\delta$ is the step size (Note that for $P_d \leq P_{\text{del,min}}$ and $r_p=\infty$, Gaussian inputs are optimal \cite{Shannon_1948} and for $P_d \leq P_{\text{del,min}}$ and $r_p<\infty$, the optimal distributions for the input amplitude $\pmb{r}$ are discrete with a finite number of mass points \cite{Shamai_BarDavid_1995}).  Set $m=1$.
  \item Utilizing interior-point algorithm, minimize the objective function in (\ref{E78}) initialized by a random guess.
  \item Once the optimal positions and their respective probabilities are found, the answer is validated by checking the average power constraint and the necessary and sufficient KKT conditions in (\ref{E70}). If the conditions are not satisfied, the initial guess is changed. We continue changing the initial guess for a large number of times.
  \item If the KKT conditions are not satisfied, the number of mass points is increased by one. We continue from stage $1$ to $4$ until at some values of $m$, KKT conditions are met.
  \item Obtain the delivered power corresponding to the optimal solution.
\end{enumerate}

\begin{figure}
\begin{centering}
\includegraphics[scale=0.27]{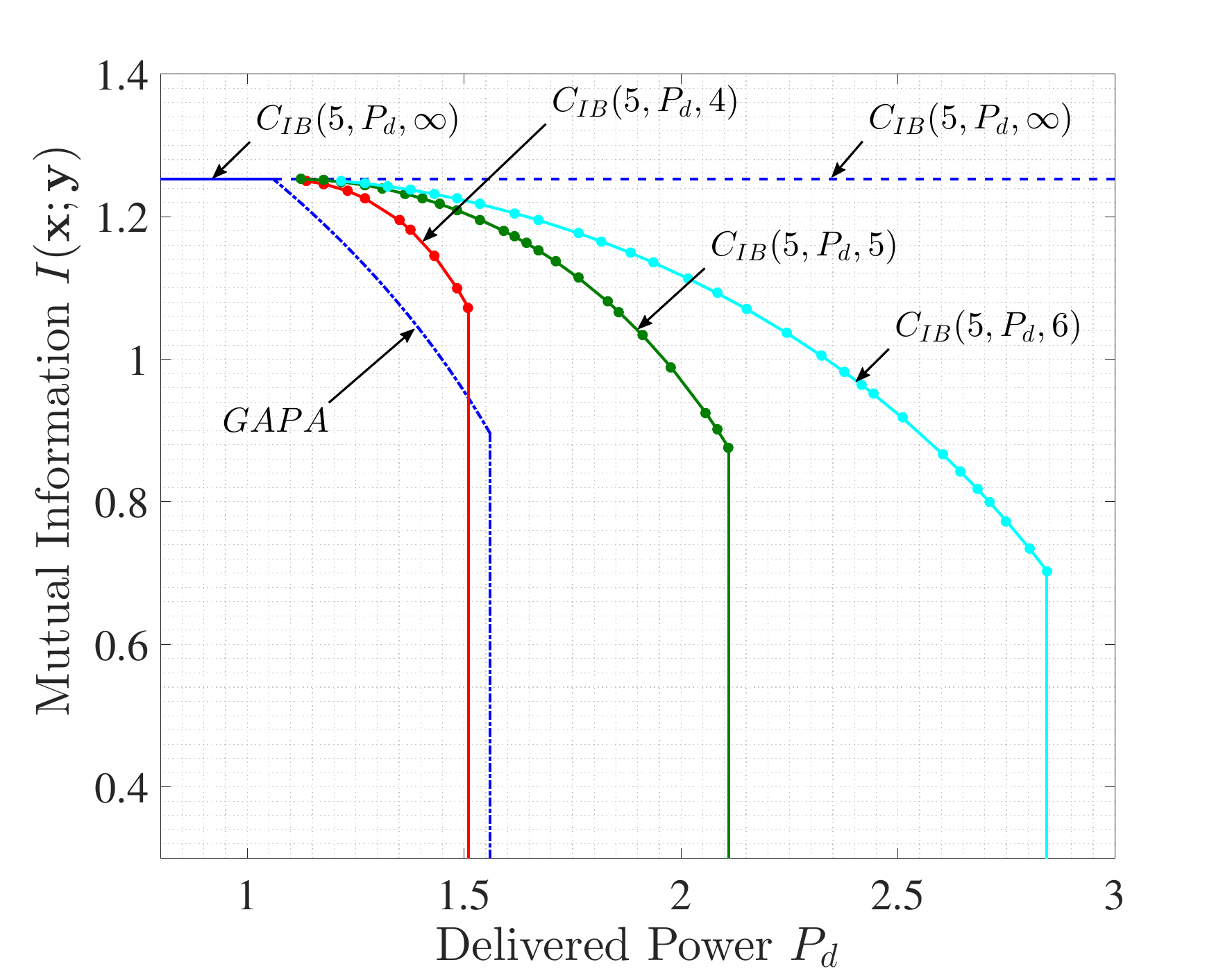}
\caption{Mutual information $I(\pmb{x};\pmb{y})$ corresponding to the complex Gaussian inputs (denoted by GAPA).  Mutual information $I(\pmb{x};\pmb{y})$ corresponding to the optimal solutions of (\ref{E76}) with respect to different values of the minimum delivered power constraint $P_d$  with amplitude constraints $r_p=4,~5,~6$ and $r_p=\infty$. Average power constraint is $P_a=5$.} \label{F1}
\par\end{centering}
\vspace{0mm}
\end{figure}

\textit{Illustration of the numerical results}:  In Figure \ref{F1}, simulation results for the transmitted information in terms of mutual information $I(\pmb{x};\pmb{y})$ and harvested power in terms of the expectation $\mathbb{E}[g_{\text{NL}}(|\pmb{x}|)]$ are illustrated for an average power constraint $P_a=5$ and $g_{\text{NL}}(r)=0.01(r^4+r^2+1)$. The horizontal solid line corresponds to the AWGN channel capacity under an average power constraint $P_a=5$, (i.e., $C_{\text{IB}}(5,P_d ,\infty)$) achieved by only a CSCG distribution. The horizontal dashed line related to $C_{\text{IB}}(5,P_d ,\infty)$ corresponds to the capacity under an average power constraint $P_a=5$, which is not achievable, however, can be approached arbitrarily (see Theorem \ref{T0}). $C_{\text{IB}}(5,P_d ,4)$, $C_{\text{IB}}(5,P_d ,5)$ and $C_{\text{IB}}(5,P_d ,6)$ correspond to the optimal solution in (\ref{E76}) for $r_p=4,~5$ and $6$, respectively. The RP region obtained from Gaussian inputs is denoted by Gaussian Asymmetric Power Allocation (GAPA)\footnote{Note that there is no amplitude constraint for this input.\label{footnote_21}}. The distributions obtained numerically by restricting the probability optimization space are denoted as numerically obtained input distributions. As it is observed from Figure \ref{F1}, numerically obtained input distributions yield significantly larger RP region compared to the region corresponding to GAPA. It is also observed that by increasing the amplitude constraint $r_p$, the RP region tends to the RP region corresponding to $r_p=\infty$. This observation is inline with Remark \ref{Rem1}, that increasing $r_p$, reduces the dependency of the capacity on $r_p$. Note that given the value of $r_p$, the amount of harvested power at the receiver is limited. This is the reason for the vertical lines corresponding to $C_{\text{IB}}(5,P_d ,4)$, $C_{\text{IB}}(5,P_d ,5)$ and $C_{\text{IB}}(5,P_d ,6)$.

\begin{figure}
\begin{centering}
\includegraphics[scale=0.27]{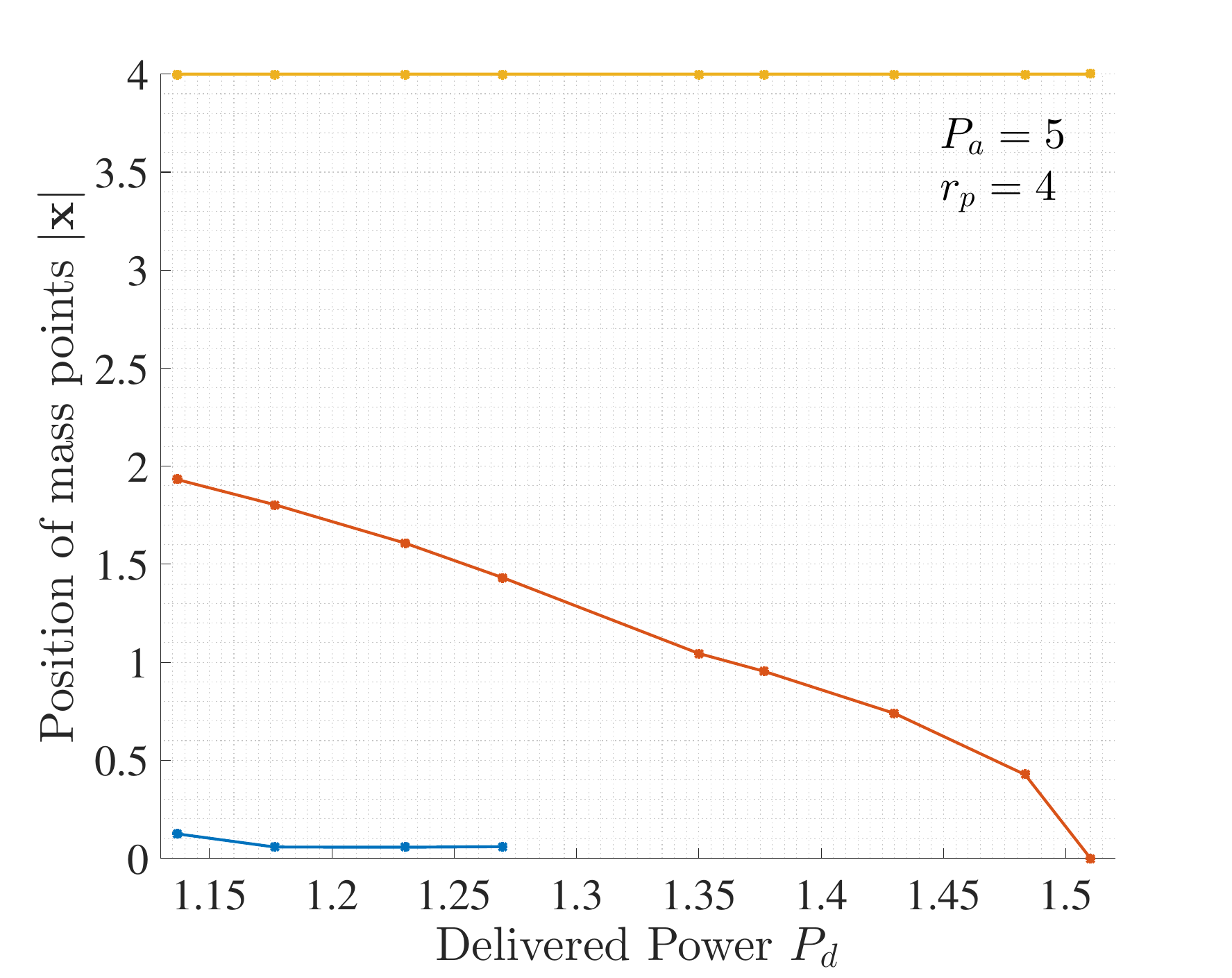}
\caption{The position of the optimal mass points for $C_{\text{IB}}(5,P_d,4)$ versus different values of the minimum delivered power $P_d$ constraint.} \label{F2}
\par\end{centering}
\vspace{0mm}
\end{figure}

\begin{figure}
\begin{centering}
\includegraphics[scale=0.27]{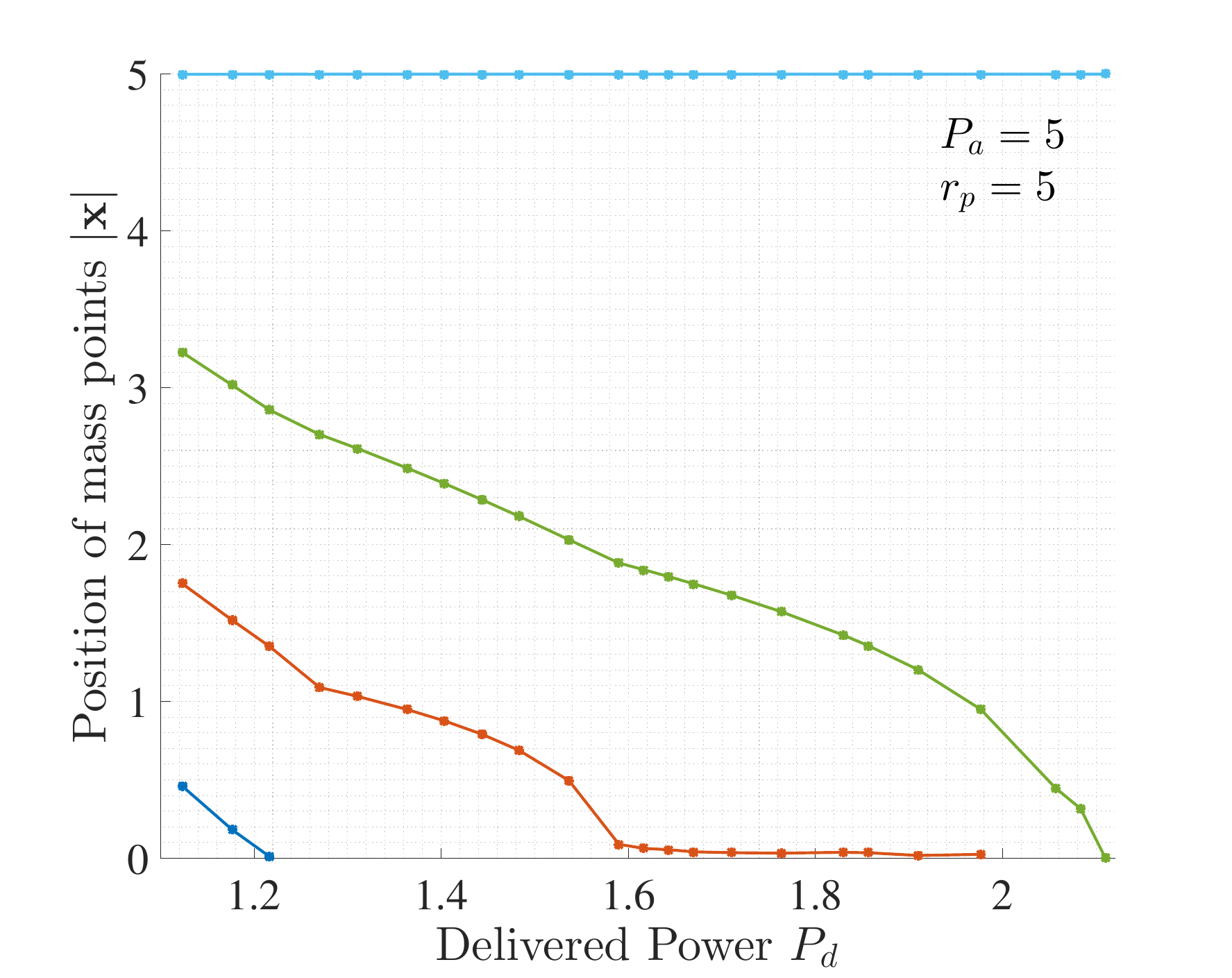}
\caption{The position of the optimal mass points for $C_{\text{IB}}(5,P_d,5)$ versus different values of the minimum delivered power $P_d$ constraint.} \label{F3}
\par\end{centering}
\vspace{0mm}
\end{figure}

\begin{figure}
\begin{centering}
\includegraphics[scale=0.27]{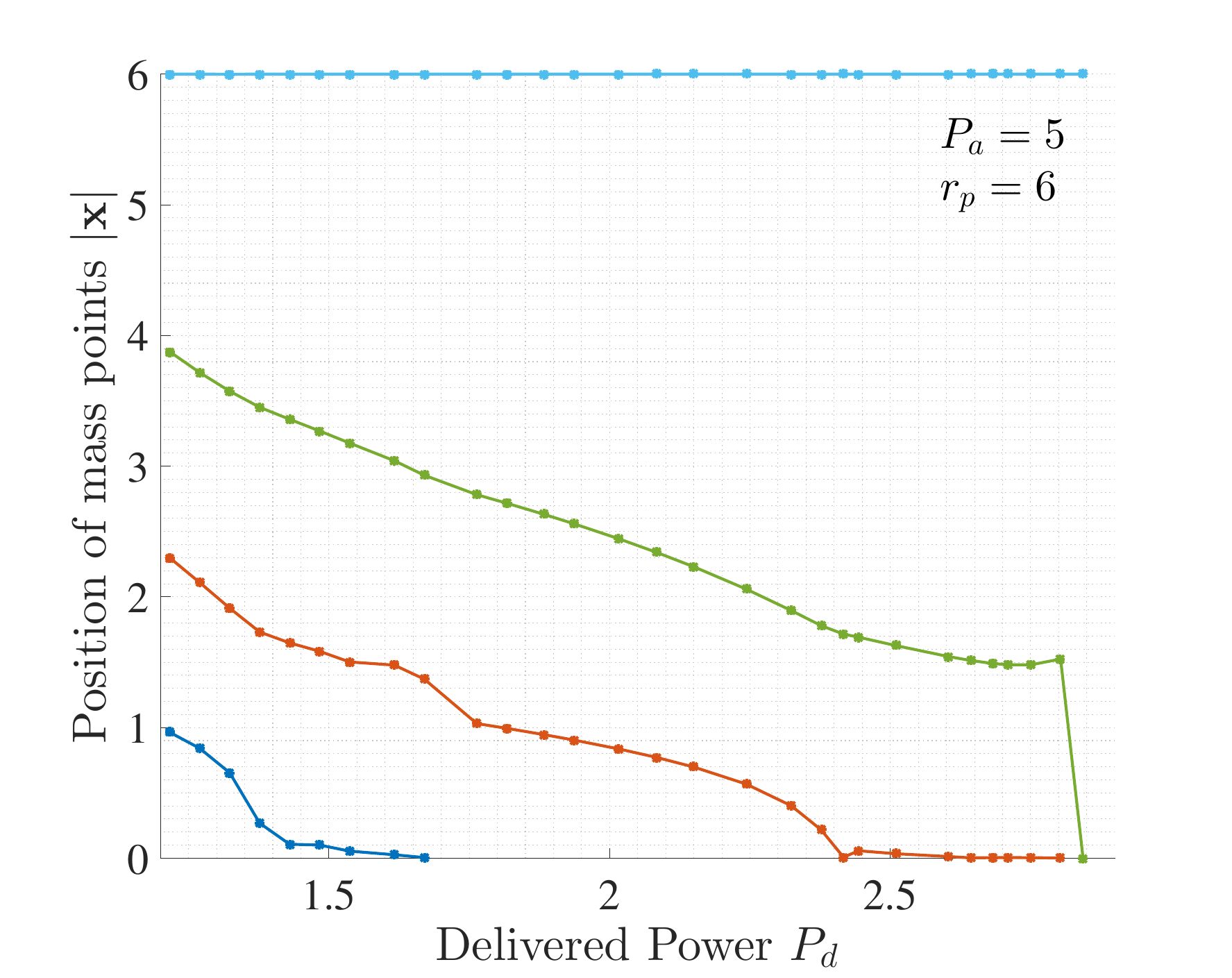}
\caption{The position of the optimal mass points for $C_{\text{IB}}(5,P_d,6)$ versus different values of the minimum delivered power $P_d$ constraint.} \label{F4}
\par\end{centering}
\vspace{0mm}
\end{figure}

In Figures \ref{F2}, \ref{F3} and \ref{F4} the position of the mass points $\pmb{r}=|\pmb{x}|$ corresponding to $C_{\text{IB}}(5,P_d ,4)$, $C_{\text{IB}}(5,P_d ,5)$ and $C_{\text{IB}}(5,P_d ,6)$ are illustrated, respectively, with respect to different delivered power constraints $P_d$. It is observed that by increasing the delivered power constraint $P_d$ at the receiver, the number of mass points decreases.  Also, as it is seen from the figures, one of the mass points is always equal to $r_p$.

\begin{figure}
\begin{centering}
\includegraphics[scale=0.27]{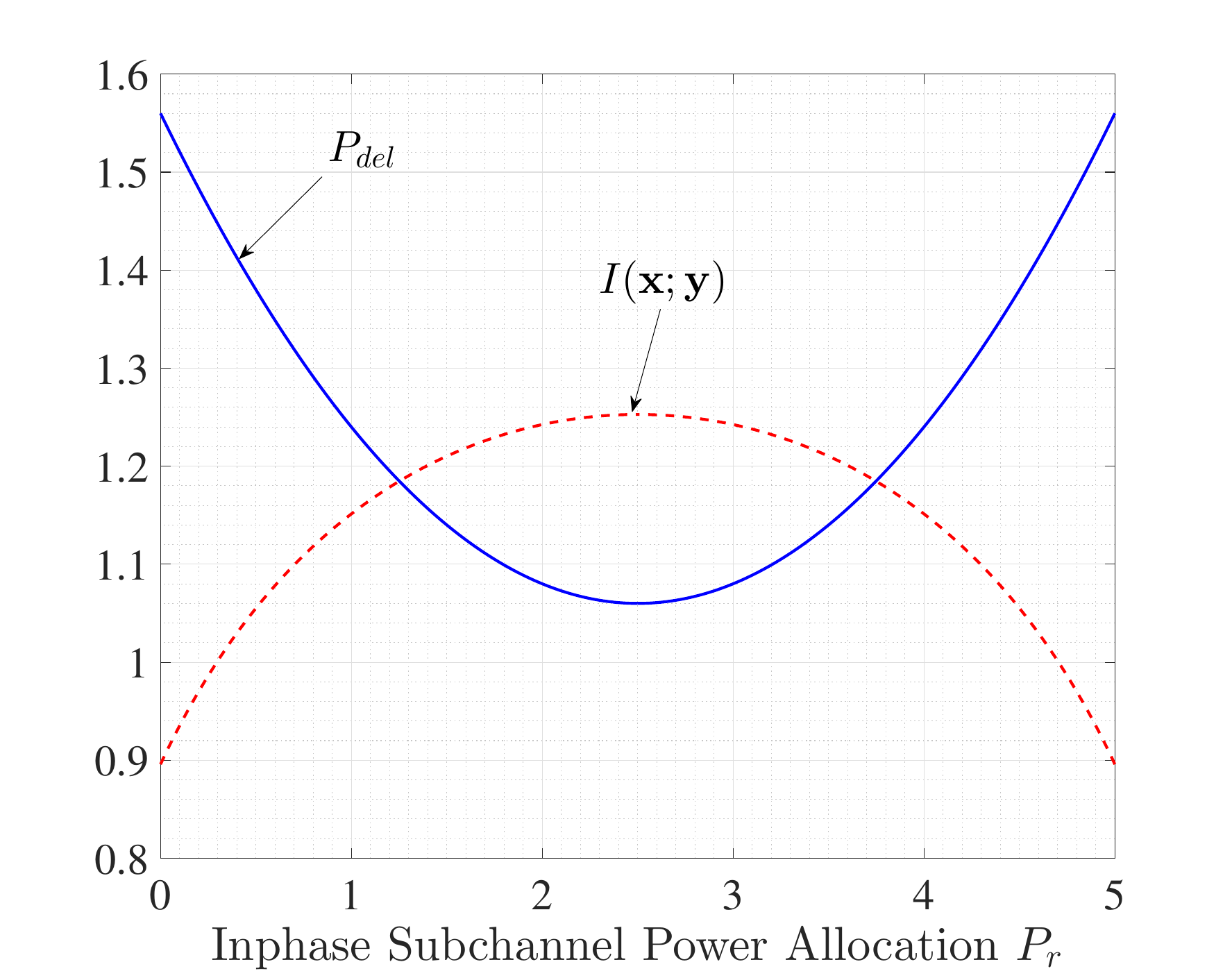}
\caption{Mutual information $I(\pmb{x};\pmb{y})$ (Red dashed line) and delivered power $P_{\text{del}}$ (blue solid line) corresponding to the complex Gaussian inputs with asymmetric power allocation. The transmitted information rate is maximized for $P_i=P_r=\frac{P_a}{2}$ and delivered power is maximized when $P_i=0,P_r=P_a$ or $P_i=P_a,P_r=0$.} \label{F11}
\par\end{centering}
\vspace{0mm}
\end{figure}

In Figure \ref{F11},  the information rate $I(\pmb{x};\pmb{y})$ and delivered power $P_{\text{del}}$ for complex Gaussian inputs is shown versus the inphase subchannel power allocation $P_i$ ($P_r=P_a-P_i$). In line with the previous results, it is observed that (unlike the linear model for the EH, that is $g(r)=\alpha_1 r^2$ in (\ref{E54})), under nonlinear model for the EH, the information rate and delivered power are maximized and minimized, respectively, for $P_i=P_r=\frac{P_a}{2}$. Alternatively the information rate and delivered power are minimized and maximized, respectively when $P_i=0,P_r=P_a$ or $P_i=P_a,P_r=0$.

Finally, we note that the algorithm used for finding numerically obtained input distributions is extremely sensitive on the first guess as the number of mass points $m$ increases. This is due to the fact that the optimization of the capacity given that the number of mass points $m$ is fixed, is not a concave function. This, accordingly, makes the problem computationally demanding with $m$.

\subsection{Realistic Circuit Simulations for WPT}\label{Kim}
\begin{figure}
\begin{centering}
\includegraphics[scale=0.24,angle =-90]{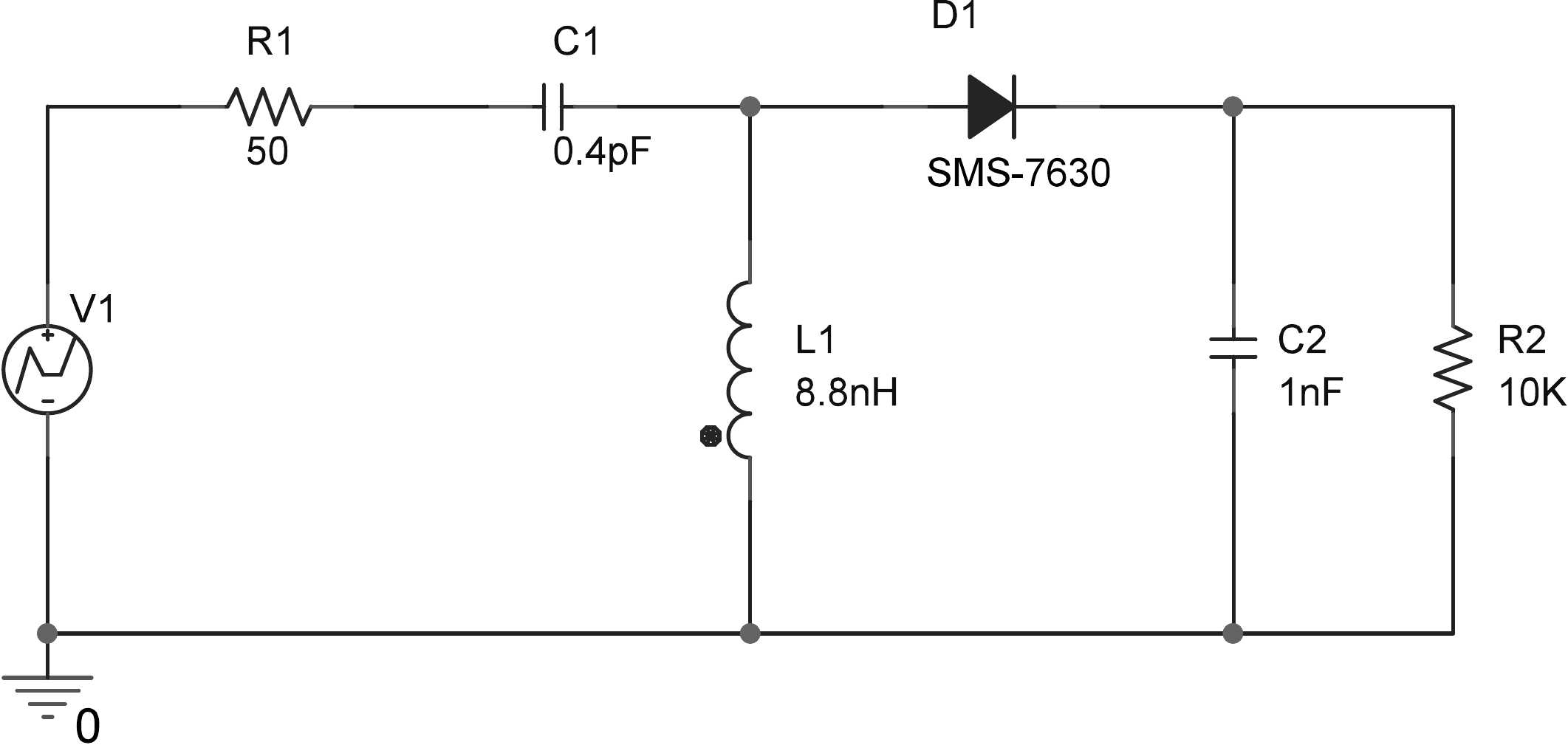}
\caption{Conventional single series rectifier circuit consisting of a rectifying diode, impedance matching circuit, and low pass filter.} \label{TD_schematic}
\par\end{centering}
\vspace{0mm}
\end{figure}

In order to assess and validate the analysis and the benefits of OOK signalling\footnote{For OOK signalling, we use the distributions introduced in (\ref{E42}) for different values of the parameter $l$.\label{footnote_22}} and asymmetric Gaussian distribution (from WPT perspective only), we designed, optimized and simulated the rectenna circuit of Figure \ref{TD_schematic}. We used a conventional single series rectifier circuit that consists of a rectifying diode, impedance matching circuit, and low pass filter. The Schottky diode Skyworks SMS7630 is chosen for the rectifying diode because it requires low biasing voltage level, which is suitable for low power rectifier. The impedance matching and low pass filter circuits are designed for an inphase 4-tone multisine input signal centered around 2.45GHz with an average power of -20dBm and with 2.5MHz inter-carrier frequency spacing. The load impedance $R2$ is chosen as $10\mathrm{K} \Omega$ in order to reach maximum RF-to-DC conversion efficiency with the 4-tone multisine waveform. The matching network capacitor $C1$, inductor $L1$ and output capacitor $C2$ values are optimized (using an iterative process) to maximize the output DC power under a given load impedance and for the given multisine input waveform at -20dBm RF input power. The chosen values are given by 0.4pF for $C1$, $8.8$nH for $L1$, and $1$nF for $C2$. The antenna impedance is set as $\mathrm{R}1 = 50 \Omega$ and the voltage source $\mathrm{V}1$ is expressed as $\mathrm{V1} = 2Y_{\mathrm{rf}}(t) \sqrt{R1}$.

In Table \ref{Tbl}, the measured delivered DC power is shown for four types of channel input, namely, continuous wave (CW)\footnote{A single tone with frequency 2.45GHz. \label{footnote_23}}, Complex Gaussian (CG), Real Gaussian (RG) and inputs of (\ref{E42}) for different values of parameter $l$ is shown. A first observation is to note that the second moment (i.e., average input power) of the input distribution is the same for all distributions, though a significant range of harvested DC power is observed. This is due to the rectenna nonlinearity that favors distributions with a large fourth moment. Indeed, the fourth moment increases proportionally to 1, 2, 3 and $l^2$ for the CW, CG, RG and OOK signalling (with $l$), respectively. This shows that the nonlinearity model through a polynomial expansion with a second and fourth order terms as in (\ref{eqn:1}) predicts the dependency of the rectenna nonlinearity on the input signal quite accurately, and confirm observations made in \cite{Clerckx_Bayguzina_2016,Clerckx_Bayguzina_2017,Clerckx_2016}. Recall that the linear model of the rectifier would not capture this dependency since it only accounts for the second order term in (\ref{eqn:1}) \cite{Clerckx_Bayguzina_2016, Clerckx_2016}. A second observation is the significantly larger power delivered with inputs in (\ref{eqn:1}) compared to other schemes. Specifically, the maximum delivered power occurs at $l=4$. The reason that the delivered power decreases for $l>4$ is due to the finite RC constant in the low pass filter of the rectenna.

\begin{table}
 \centering
\begin{tabular}{ |c|c|c|c| }
\hline
Transmission type & Delivered DC Power ($\mu$W) \\
\hline
\multirow{3}{4em}{CW \\CG\\ RG \\l=2\\l=3\\l=4\\l=5} & 1.0959  \\
& 1.5296 \\
& 1.7547  \\
& 2.6899  \\
& 3.4262  \\
& 3.4884  \\
& 3.2965  \\
\hline
\end{tabular}
\caption{Delivered power for Continuous Wave (CW), Complex Gaussian (CG), Real Gaussian (RG) and OOK signalling with different pdfs.}\label{Tbl}
\end{table}

\section{Discussion and Future Works}\label{sec_Disc}
In the following, we discuss a number of interesting research avenues that can be considered in the future.

\begin{itemize}
  \item Note that the delivered power in (\ref{eqn:22}), contains odd moments of the channel input $\pmb{x}$. Accordingly, for the problem considered in (\ref{E3}), it is interesting to find optimal input distributions when the function $g(r)$ (we recall that $g(r)$ models the baseband representation) contains odd powers of the argument.

  \item The practical EHs exhibit nonlinear behaviors since their efficiency becomes different (not constant) when the received RF power level changes. Specifically, the efficiency is very small in low RF power level (due to the turn-on voltage of the diode), is large in the middle RF power level, and is again very small in the high RF power level (due to the reverse breakdown of the diode). In order to capture this behaviour, the function $g(r)$ should not tend to infinity when $r\rightarrow\infty$. Accordingly, finding optimal inputs for bounded $g(r)$ is of interest.

  \item The problem considered in (\ref{E3}), is indeed an optimization over circular symmetric solutions. However, in practical SWIPT problems, harvesters are also phase dependent and circuit simulations reveal that phase variations in the channel input can also affect the delivered power at the receiver significantly \cite{Bayguzina_Clerckx_1802.06512}. Hence, it is interesting to develop a systematic approach in order to capture the effect of phase variations as well.

  \item Note that the harvester's input is the RF signal $Y_{\text{rf}}(t)$ (see (\ref{eqn:1})), and therefore, in the baseband representation (for nonlinear harvesters), it appears that we have higher order moment of the baseband equivalent of the channel output, i.e., $Y(t)$ (see (\ref{eqn:5}) in Appendix \ref{app:1}). Accordingly, to represent the signal perfectly in terms of its samples, we require to consider more values of the baseband channel output $Y(t)$ between any consecutive information samples (see (\ref{E77}) in Appendix \ref{app:1}). If we allow correlation among the samples in Section \ref{Sec_Application} (recall that iid assumption on samples $\pmb{x}_k$ for different $k$ in (\ref{E90}) is an imposed condition), the problem becomes cumbersome to analyze. However, it seems to the authors that from a power harvesting point of view, correlation among different samples is good, in opposition to information transmission. Hence, it is also interesting to consider even very simple achievable schemes which utilize the effect of correlation.

  \item Finally, we note that the results presented here can be extended to vector Gaussian channels with bounded inputs \cite{Rassouli_Clerckx_16} and Gaussian multiple access channels \cite{Rassouli_Varasteh_Gunduz_17}, utilizing the similar tools presented therein.

\item There might be interesting connections to make with other systems subject to nonlinear responses. In optical communications, for instance, the nonlinearity is commonly compensated and transmission is performed using constellations approximating the zero-mean Gaussian distribution optimum for AWGN channels (e.g. ring constellations) \cite{Essiambre_Kramer_Winzer_Foschini_Goebel}. The information theoretic limits of optical channels are studied by modelling the nonlinear optical communication channel as a linear channel with a multiplicative noise or using a finite-memory model with additive noise \cite{Essiambre_Kramer_Winzer_Foschini_Goebel,Agrell_Alvarado_Durisi_Karlsson}. On the contrary, in SWIPT, the diode nonlinearity is exploited in the signal design and in the characterization of the RP region, therefore leading to non-zero mean Gaussian inputs and enlarged region compared to that obtained with zero-mean inputs.

 \item As the average power constraint grows, obtaining the optimal distribution gets computationally demanding. One alternative to reduce the computational load of the optimization is to follow the approach presented in \cite{Huang_Meyn}, where a modified cutting-plane algorithm is applied on a piecewise-linear approximation of mutual information.

\end{itemize}

\section{Conclusions}\label{Sec_Conclusion}
In this paper, we studied the capacity of a complex AWGN channel under transmit average power, amplitude and receiver delivered power constraints. We focused on nonlinear delivered power constraints at the receiver. We showed that under an average power constraint and for any given delivered power constraint, the capacity of an AWGN channel can be either achieved or approached arbitrarily. In line with the similar results in the literature, we showed that including the amplitude constraint causes the optimal inputs to be discrete with a finite number of mass points. As an application of the presented results, we considered SWIPT over a complex AWGN channel in the presence of a nonlinear EH at the receiver. Defining the RP region, we provided two inner bounds for the RP region. Considering general complex Gaussian inputs as the first inner bound, we showed that the optimal Gaussian inputs are zero mean. A tradeoff between the transmitted information and delivered power is recognized by allocating the power budget asymmetrically between the real and imaginary subchannels. By restricting the optimization probability space, we utilized the obtained results in this paper to derive the second inner bound. Numerical results reveal that there are significant improvements in the second inner bound with respect to the first inner bound corresponding to complex Gaussian inputs.

\section{Acknowledgment}
The authors would like to thank Junghoon Kim for providing the circuit simulation results in Section \ref{Kim}. We also would like to express our sincere gratitude towards the anonymous Reviewers of this manuscript for providing insightful comments.

\section{Guide on Using Appendices}\label{Guide}
In the following, the proof of the main results of the paper, i.e., Theorem \ref{T0}, Theorem \ref{T1}, Lemma \ref{Prop1} and Lemma \ref{Prop2} are provided in Appendices \ref{A0}, \ref{A1}, \ref{app:1} and \ref{app:2}, respectively. The preliminary results required for the main proofs are provided in Appendix \ref{First_app} where the proofs of some of them are provided in Appendices \ref{app:3}, \ref{A6}, \ref{A8}, \ref{A4}, \ref{A5}, \ref{A3}, \ref{app:3}.

\appendices

\section{Lemmas}\label{First_app}
In this appendix, we provide the lemmas required to prove Theorems \ref{T0} and \ref{T1}.

\begin{lem}\label{Lem_single}
The capacity of the channel in (\ref{E69}) with input subject to
\begin{align}
    \left\{\!\!\!\!\!\!\!\begin{array}{lll}
    &\frac{1}{n}\sum_{k=1}^{n}|\pmb{x}_k(m)|^2  \leq  P_a\\
    &\frac{1}{n}\sum_{k=1}^{n}g(|\pmb{x}_k(m)|)\geq P_d   \\
    &|\pmb{x}_k(m)|\leq r_p
    \end{array}\right.,\forall m \in [1:2^{nR}],
\end{align}
is the same as the capacity of the channel in (\ref{E69}) with input samples subject to (\ref{E1}).
\end{lem}
\textit{Proof}: We follow \cite[Section 3.3]{elGamal:book}, however with the following modification. Define
    \begin{equation}
        \begin{aligned}
        C(P_a,P_d,r_p)=& \sup\limits_{f_{\pmb{x}}(x)}
        & & I(\pmb{x};\pmb{y}) \\
        & \text{s.t.}
        & & \left\{\begin{array}{lll}
        \mathbb{E}[|\pmb{x}|^2]  \leq  P_a \\
        \mathbb{E}[g(|\pmb{x}|)] \geq P_d \\
        |\pmb{x}|\leq r_p
        \end{array}\right..
        \end{aligned}
    \end{equation}

    First, we note that
    \begin{itemize}
        \item $C(P_a,P_d,r_p)$ is concave in $(P_a,P_d)$.
        \item $C(P_a,P_d,r_p)$ is nondecreasing in $P_a$ and nonincreasing in $P_d$.
        \item The continuity follows due to concavity.
    \end{itemize}

    \textbf{Proof of achievability}: In contrast to \cite[Theorem 3.2]{elGamal:book}, we are not dealing with a maximum. Let $f^\delta(x)$ be such that $I(\pmb{x};\pmb{y})$ under $f^\delta(x)$ falls within $\delta-$neighbourhood of $C(P_a/(1+\epsilon_1),P_d/(1-\epsilon_2), r_p)$ with arbitrary small $\epsilon_1$ and $\epsilon_2$. Also assume there is an $x_0$, such that, $|x_0|^2<P_a,~g(|x_0|)\geq P_d$ and $|x_0|\leq r_p$.

    In an iid fashion, generate $2^{nR}$ $n$-length sequences $x^n(m)$ according to $\Pi_{i=1}^{n}f^\delta(x_i)$.
    \begin{itemize}
        \item If $\pmb{x}^n(m)\in \mathcal{T}_{\epsilon}^{(n)}$, then transmit it, and accordingly, due to typical average lemma we have
            \begin{align}
            \left\{\!\!\!\!\!\!\!\begin{array}{lll}
            &\frac{1}{n}\sum_{k=1}^{n}|\pmb{x}_k(m)|^2  \leq  P_a\\
            &\frac{1}{n}\sum_{k=1}^{n}g(|\pmb{x}_k(m)|)\geq P_d   \\
            &|\pmb{x}_k(m)|\leq r_p
            \end{array}\right.,\forall m \in [1:2^{nR}].
            \end{align}
            Note that the last inequality is already guaranteed.
        \item If $\pmb{x}^n(m)\notin  \mathcal{T}_{\epsilon}^{(n)}$, then transmit a randomly chosen typical sequence.
    \end{itemize}

    We have now generated a coding scheme that satisfies the constraints. Analysis of error events $\textrm{Pr}(\mathcal{E}_1)$, $\textrm{Pr}(\mathcal{E}_2)$ follows along the similar lines as in \cite[Section 3.3]{elGamal:book}. Therefore, every rate $R<C(P_a/(1+\epsilon_1),P_d/(1-\epsilon_2), r_p)-\delta$ is achievable. By continuity of $C(\cdot,\cdot,r_p)$, let $\epsilon_1,\epsilon_2 \rightarrow 0$, and therefore, any rate $R<C(P_a,P_d, r_p)-\delta$ is achievable. Finally, since $\delta$ was chosen arbitrarily, any $R<C(P_a,P_d, r_p)$ is achievable.

    \textbf{Proof of converse}:
    \begin{align}\label{Br1}
    nR &\leq \sum_{k=1}^{n} I(\pmb{x}_k;\pmb{y}_k)\!+\!n\epsilon_n\\\label{Br2}
    &\leq \sum_{k=1}^{n} C(\mathbb{E}[|\pmb{x}_k|^2],\mathbb{E}[g(|\pmb{x}_k|)],B_3)\!+\!n\epsilon_n\\\label{Br3}
    &\leq nC\!\left(\!\frac{1}{n}\sum_{k=1}^{n} \mathbb{E}[|\pmb{x}_k|^2], \frac{1}{n}\sum_{k=1}^{n}\mathbb{E}[g(|\pmb{x}_k|)], B_3\!\right)\!+\!n\epsilon_n\\\label{Br4}
    &\leq nC(P_a,P_d, r_p)\!+\!n\epsilon_n,
    \end{align}
    where (\ref{Br1}) is due to Fano and Data processing inequality, (\ref{Br2}) is due to the definition, (\ref{Br3}) is due to the concavity of with respect to $P_a,~P_d$ and (\ref{Br4}) is due to monotonicity.\qed

\begin{lem}\label{lem1}
In the Levy's metric\footnote{For two cumulative distribution function $F,~G:~\mathbb{R}\rightarrow [0,1]$, the Levy's distance $L(F,G)$ is defined as $L(F,G)=\inf \{\epsilon >0|F(x-\epsilon)-\epsilon \leq G(x) \leq F(x+\epsilon)+\epsilon , \forall x\in \mathbb{R}\}$.\label{footnote_24}}, the space $\Omega_1\cap\Omega_2$ is convex, however, compact if $r_p<\infty$.
\end{lem}
\textit{Proof}: The proof is obtained by following exactly the same approach used in \cite{Smith:IC:71}. In the following, we bring a counterexample which proves that the space $\Omega_1\cap\Omega_2$ for $r_p=\infty$ is not compact. For simplicity, assume $g(r)=r^4$ (the following argument can be extended to the general definition of $g(r)$ in (\ref{E54})) and consider the following sequence of probability distributions
\begin{align}\label{E94}
    F_{\pmb{r},l}(r)=\left\{\begin{array}{cc}
                            0 & r<0,\\
                            1-\frac{1}{l^4} & 0\leq r < \sqrt[4]{P_{d}}l,\\
                           1 & r\geq \sqrt[4]{P_{d}}l,
                         \end{array}
    \right. ~l=0,1,\ldots.
\end{align}
It can be verified that $\mathbb{E}[\pmb{r}^4]= P_d$ and for integer $l\geq \sqrt[4]{P_{d}/P_a^2}$ we have $\mathbb{E}[\pmb{r}^2]\leq P_a$. However, the limiting distribution (when $l\rightarrow\infty$) is $F_{\pmb{r}}^{*}(r)=U(r)$ does not satisfy the second constraint, i.e., $\mathbb{E}[\pmb{r}^4]= 0$. This establishes that the space $\Omega_1\cap\Omega_2$ for  $P_{d},P_a<\infty$ and $r_P=\infty$, is not compact\footnote{Note that compactness is a sufficient condition for continuous functions to achieve their supremum or infimum, however, not necessary.\label{footnote_25}}.\qed

\begin{lem}\label{lem6}
For all $x\geq 0$ we have
\begin{align}\label{E37}
  I_0(x)<\inf\limits_{0< a<1} e^x\left(\frac{\hat{a}(1-e^{-2ax})}{\pi x}+\frac{ \mathrm{erf}\left(\sqrt{2ax}\right)}{\sqrt{2\pi x}}+e^{-2ax}\right),
\end{align}
where $\hat{a}=\frac{\frac{1}{\sqrt{1-a}}-1}{2\sqrt{a}}$. Furthermore, it is verified that
\begin{align}\nonumber
  &\lim\limits_{x\rightarrow 0} e^x\left(\frac{\hat{a}(1-e^{-2ax})}{\pi x}+\frac{ \mathrm{erf}\left(\sqrt{2ax}\right)}{\sqrt{2\pi x}}+e^{-2ax}\right)\\
  &\quad\quad=1+\frac{\sqrt{a}}{\pi}+\frac{\sqrt{a}}{\pi\sqrt{1-a}}.
\end{align}
Substituting $a=1/2$ in (\ref{E37}) and noting that $\mathrm{erf}(x)\leq 1$ and $1-e^{-x}\leq \sqrt{\pi x}$ we have
\begin{align}\label{E46}
  I_0(x)&<\frac{e^x}{\sqrt{\pi x}}+1.
\end{align}
Using the inequality $\sqrt{\pi x}< e^x (\sqrt{\pi}-1)$, we can further upper bound (\ref{E46}) as
\begin{align}\label{E38}
  I_0(x)&<\frac{e^x}{\sqrt{x}}.
\end{align}
\end{lem}
\textit{Proof}: See Appendix \ref{A6}.

\begin{lem}\label{lem8}
In the following integral transform
\begin{align}\label{E55}
\int\limits_{0}^{\infty} K(R,r)G(R)dR&=g(r),
\end{align}
where $g(r)$ is defined in (\ref{E54}), if $G(R)$ is restricted to be a polynomial with a finite degree, i.e., of the $\sum\limits_{i=0}^{d}c_iR^{2i}$, there are unique $c_i$'s that satisfy (\ref{E55}).
\end{lem}
\textit{Proof}: See Appendix \ref{A8}.

\begin{lem}\label{lem2}
$f_{\pmb{R}}(R;F_{\pmb{r}}),~R\geq0,~F_{\pmb{r}}\in \Omega_1\cap\Omega_2$ is bounded and continuous in both of its arguments.
\end{lem}
\textit{Proof}: Continuity of $K(R,r)$ follows by the continuity of $I_0(rR)$. Note that
\begin{align}
  K(0,r) & =K(\infty,r) =K(R,\infty) =0,\\
  K(R,0)& =Re^{-\frac{R^2}{2}}<\infty, \\\label{E20}
  K(\infty,\infty)& <\sqrt{\frac{R}{r}}e^{-\frac{(R-r)^2}{2}}<\infty,
\end{align}
where (\ref{E20}) is due to (\ref{E38}). Therefore the function $K(R,r)$ is bounded. After some algebraic manipulations (details have been provided in Appendix \ref{App_bounding}), it can be shown that
\begin{align}\label{E26}
K(R,r)<1.
\end{align}
Continuity of $f_{\pmb{R}}(R;F_{\pmb{r}})$ is obtained by following the same steps as in \cite[Lemma 3]{Tchamkerten_2004}. From (\ref{E26}) and $K(R,r)>0$ it can also be easily verified that
\begin{align}\label{E31}
0< f_{\pmb{R}}(R;F_{\pmb{r}})< 1,~R>0.
\end{align}\qed

The following Lemma is indeed a generalization of \cite[Theorem 13]{Fahs_AbouFaycal_2012} to complex channels.
\begin{lem}\label{lem9}
Let $\pmb{n}$ be a CSCG random variable of variance $2$, and let $\pmb{x}$ be a complex random
variable that is independent of $\pmb{n}$. The PDF $f_{\pmb{y}}(\cdot)$ of the random variable $\pmb{y}=\pmb{x}+\pmb{n}=|\pmb{y}|e^{j\pmb{\theta}}$ decays as $\exp{-|y|^2/2}$ or slower, that is
\begin{align}\label{tag1}
f_{\pmb{y}}(y)\neq \mathcal{O}\left(e^{-A|y|^2}\right),~\forall A>\frac{1}{2}.
\end{align}
\end{lem}
\textit{Proof}: By calculating the characteristic function of the complex random variable $\pmb{y}$, we have
\begin{align}
  &|M_{\pmb{y}}(z=r_ze^{j\theta_z})|=|\mathbb{E}[e^{j\text{Re}(z^{*}\pmb{y})}]|\\
  &\quad\quad=|\mathbb{E}[e^{j\text{Re}(z^{*}\pmb{x})}]|\cdot|\mathbb{E}e^{j\text{Re}(z^{*}\pmb{n})}]|\\
  &\quad\quad\leq|\mathbb{E}e^{j\text{Re}(z^{*}\pmb{n})}]|\\
  &\quad\quad=\Bigg|\int\limits_{0}^{\infty}\int\limits_{0}^{2\pi}\frac{r_n}{2\pi}e^{-\frac{r_n^2}{2}}e^{jr_nr_z\cos(\theta_n- \theta_z)}dr_nd\theta_n\Bigg|\\
  &\quad\quad=\Bigg|\int\limits_{0}^{\infty}r_ne^{-\frac{r_n^2}{2}}I_0(jr_nr_z)dr_n\Bigg|\\\label{E63}
  &\quad\quad=e^{-\frac{r_z^2}{2}},
\end{align}
where (\ref{E63}) is due to the transform $t=\frac{r_n^2}{2}$ and \cite[ET I 197(20)a]{gradshteyn2007}.

Continuity of $\pmb{y}$ is verified due to continuity of the complex Gaussian noise $\pmb{n}$. From Lemma \ref{lem2}, existence of the pdf of $\pmb{y}$ is guaranteed. Hence, the result of the lemma is proved by Hardy's theorem (see \cite{Hardy}) and (\ref{E63}) and noting that any pdf in the form of $f_{\pmb{y}}(y)=\mathcal{O}(e^{-A|y|^2}),~A>1/2$ is identically zero, i.e., $f_{\pmb{y}}(y)=0$, which is not a legitimate pdf.
\qed.

\begin{lem}\label{lem4}
$f_{\pmb{R}}(R;F_{\pmb{r},n})\ln f_{\pmb{R}}(R;F_{\pmb{r},n})$ for $R\geq0$, $F_{\pmb{r},n}\in \Omega_1\cap\Omega_2$ is dominated by the following absolutely integrable function
\begin{align}
g(R)=\left\{\begin{array}{cc}
     4 & R\leq 2\\
     \frac{c}{R^{\frac{3}{2}}} & R>2
   \end{array}\right.,
\end{align}
where $c=4(128+4P_a)^{\frac{3}{4}}$.
\end{lem}
\textit{Proof}: See Appendix \ref{A4}.

\begin{lem}\label{lem5}
For every $F_{\pmb{r}}\in \Omega_1\cap\Omega_2$, $H(F_{\pmb{r}})$ exists, and is continuous, strictly concave and weakly differentiable.
\end{lem}
\textit{Proof}: See Appendix \ref{A5}.

The following series will be useful throughout the proof of Lemma \ref{Prop1}.
\begin{lem}\label{Lemma1} Recalling that $a_l=\text{sinc}(l+1/2)$ for integer $l$, we have the following series\footnote{The summations are from $-\infty$ to $\infty$. They are removed due to brevity.\label{footnote_26}}:
\begin{align}
S_0&\triangleq \sum_{l} a_l^2=1,\\
S_1&\triangleq \sum_{l}\sum_{k:k\neq l} a_l  a_k=0,\\
S_2&\triangleq \sum_{l}\sum_{k:k\neq l}\sum_{\substack{d:d\neq l\\d\neq k}}\sum_{\substack{m:m\neq l\\m\neq d\\m\neq k}} a_l  a_k a_d a_m=0,\\
S_3&\triangleq\sum_{l}\sum_{k:k\neq l} a_l^2  a_k^2=\frac{2}{3},\\
S_4&\triangleq\sum_{l}\sum_{k:k\neq l}\sum_{\substack{d:d\neq l\\d\neq k}} a_l^2  a_k a_d=-\frac{1}{3},\\
S_5&\triangleq\sum_{l} a_l^4=\frac{1}{3},\\
S_6&\triangleq\sum_{l}\sum_{k:k\neq l}a_l^3  a_k=\frac{1}{6}.
\end{align}
\end{lem}
\textit{Proof}: See Appendix \ref{app:3}.

\section{Proof of Theorem \ref{T0}}\label{A0}
It is easy to verify that for a given average power constraint $P_a$, capacity $C(P_a,P_d,\infty)$ is a non-increasing function with $P_d$. Therefore, we have
\begin{align}
C(P_a,0,\infty)\geq C(P_a,P_d,\infty).
\end{align}
Note that $C(P_a,0,\infty)=\ln(1+P_a/2)$ and is achieved by a unique CSCG input distribution as $\pmb{x}\sim \mathcal{CN}(0,P_a)$ (with its amplitude $\pmb{r}$ distributed as Rayleigh distribution according to the CDF $F_{\pmb{r}_R}(r)=1-e^{-\frac{r^2}{2P_a}}$).\footnote{The subscript $R$ stands for the Rayleigh distribution.\label{footnote_27}} The uniqueness of the input is proved by noting that the integral transform (\ref{E93}) is invertible (the proof follows exactly the same steps in \cite[appendix II]{Shamai_BarDavid_1995}. Accordingly, we have removed the proof for brevity.) The delivered power corresponding to $\pmb{x}\sim \mathcal{CN}(0,P_a)$ is obtained as
\begin{align}
P_G=\frac{1}{P_a}\int\limits_{0}^{\infty}r g(r) e^{-\frac{r^2}{2P_a}}dr.
\end{align}
Hence, we have
\begin{align}\label{E5}
C(P_a,0,\infty)= C(P_a,P_d,\infty),~P_d\leq P_G.
\end{align}

Since $\pmb{x}\sim \mathcal{CN}(0,P_a)$ is the only distribution achieving the capacity $C(P_a,0,\infty)$, therefore, $C(P_a,0,\infty)$ is not achieved for $P_d> P_G$\footnote{We note that the reason the capacity is not achieved is not due to the fact that the optimization probability space is open (i.e., is not compact), but rather is due to the contradiction in uniqueness of the achievable input.\label{footnote_28}}. In what follows, we show that, (\ref{E5}) holds for $P_d> P_G$. In other words, when $P_d> P_G$, any rate lower than $C(P_a,0,\infty)$ can be achieved by a distribution whose corresponding delivered power is equal to $P_d$. Consider the following sequence of distribution functions
\begin{align}\label{E42}
F_{\pmb{r}_l}(r)=\left\{\begin{array}{ll}
                 0&r<0\\
                 1-\frac{1}{l^2}   & 0\leq r<\sqrt{P_a}l \\
                 1 & r\geq  \sqrt{P_a}l
               \end{array}\right.,~l=2,3,\ldots.
\end{align}
It is easy to verify that $F_{\pmb{r}_l}(r),~l=2,\ldots$ satisfy $\mathbb{E}_{F_{\pmb{r}_l}}[\pmb{r}_l^2]=P_a$, hence, satisfying the average power constraint. Also, for the delivered power constraint we have
\begin{align}
P_{d,l}\triangleq\mathbb{E}_{F_{\pmb{r}_l}}[g(\pmb{r}_l)]=\alpha_0+ \alpha_1 P_a + \sum\limits_{i=2}^{n}\alpha_i P_a^i l^{2i-2}.
\end{align}
Since $n\geq 2$ by construction, it is guaranteed that there exists an integer number $L$, such that for $l>L$, $P_{d,l}\geq P_d$ (note that $P_{d,l}\rightarrow\infty$ as $l\rightarrow\infty$). Due to Lemma \ref{lem1}, time sharing is valid in our system model. Hence, we can construct a complex input with its phase uniformly distributed over $[-\pi,\pi)$ and its amplitude distributed according to the following CDF
\begin{align}\label{E80}
F_{\pmb{r}_{ts}}(r)=(1-\tau) F_{\pmb{r}_R}(r)+\tau F_{\pmb{r}_l}(r),~\tau\in(0,1),~l>L,
\end{align}
where the subscript $ts$ in $F_{\pmb{r}_{ts}}$ stands for time-sharing. By choosing $\tau=(P_d-P_G)/(P_{d,l}-P_G)$, we have $0<\tau<1$ and  the constraints
\begin{align}
\left\{\begin{array}{l}
                 \mathbb{E}_{F_{\pmb{r}_{ts}}}[\pmb{r}_{ts}^2]=P_a,\\
                \mathbb{E}_{F_{\pmb{r}_{ts}}}[g(\pmb{r}_{ts})]= P_d.
               \end{array}\right.
\end{align}
are both satisfied. Accordingly, we have
\begin{align}\nonumber
H(F_{\pmb{r}_R})&>H(F_{\pmb{r}_{ts}})\\\nonumber
&> (1-\tau) H(F_{\pmb{r}_R})+\tau H(F_{\pmb{r}_l})\\\label{E79}
&>-2,~\tau\in(0,1),~l>L,
\end{align}
where the first inequality in (\ref{E79}) is due to non-increasing behaviour of $C(P_a,P_d,\infty)$ and uniqueness of $\pmb{x}\sim \mathcal{CN}(0,P_a)$ in achieving $C(P_a,0,\infty)$. The second inequality in (\ref{E79}) is due to strict concavity of the entropy $H(F_{\pmb{r}})$ (see Lemma \ref{lem5}), and the third inequality is verified by noting that $H(F_{\pmb{r}})=\int_{0}^{r_p}h(r;F_{\pmb{r}})dF_{\pmb{r}}(r)$ and $h(r;F_{\pmb{r}})\geq -2$ for any $F_{\pmb{r}}\in \Omega_1\cap \Omega_2$ (see (\ref{E59}) in Appendix \ref{A1}). For a given $P_d$, we can increase $l$ arbitrarily. Therefore $\tau$ can be made arbitrarily close to zero by letting $l\rightarrow\infty$. Therefore, by letting $\tau$ tend to zero (equivalently letting $P_{d,l}\rightarrow\infty$), the result of Theorem \ref{T0} is concluded. We note that there is no distribution achieving the supremum.

\section{Proof of Theorem \ref{T1}}\label{A1}

The main steps of the proof of Theorem \ref{T1} are parallel to those provided in \cite{Shamai_BarDavid_1995,Smith:IC:71,AbouFaycal_Trott_Shamai_2001,Tchamkerten_2004,Fahs_AbouFaycal_2012}. Therefore, we provide the details for the different arguments and briefly mention (for brevity) the straightforward outcomes.

Since the set $\Omega_1\cap\Omega_2$ is compact  for $r_p<\infty$ (see Lemma \ref{lem1}) and $H(F_{\pmb{r}})$ is continuous (see Lemma \ref{lem4}), it is verified that the supremum in (\ref{E3}) is achieved and therefore it can be replaced by maximum. Due to convexity of the set $\Omega_1\cap\Omega_2$ (see Lemma \ref{lem1}) and strict concavity of $H(F_{\pmb{r}})$ (see Lemma \ref{lem4}), it is concluded that the maximum is achieved by a unique $F_{\pmb{r}^o}\in\Omega_1\cap\Omega_2$. It is verified from Lemmas \ref{lem1} and \ref{lem4} that the conditions of the Lagrangian theorem \cite[Section 8.3]{Luenberger_1969} are met. By writing the Lagrangian we have
\begin{align}\nonumber
&L(F_{\pmb{r}},\lambda,\mu)\\\label{E48}
&=\int\limits_{0}^{r_p}h(r;F_{\pmb{r}})-\lambda (r^2-P_a)+\mu(g(r)-P_d)dF_{\pmb{r}}(r),
\end{align}
where $\lambda\geq 0,~\mu\geq0$ are Lagrange multipliers and $h(r;F_{\pmb{r}})$ is defined in (\ref{E14}). By weak differentiability of $H(F_{\pmb{r}})$ (see Lemma \ref{lem4}) and the linear constraints in (\ref{E47}), the weak derivative \cite[Section 7.4]{Luenberger_1969} of (\ref{E48}) with respect to $F_{\pmb{r}^o}$ reads as
\begin{align}\nonumber
&L^{'}_{F_{\pmb{r}^o}}(F_{\pmb{r}},\lambda,\mu)\\
&=\int\limits_{0}^{r_p}h(r;F_{\pmb{r}^o})-\lambda r^2 +\mu g(r) -K_{F_{\pmb{r}^o}} dF_{\pmb{r}}(r),
\end{align}
where $K_{F_{\pmb{r}^o}}\triangleq H(F_{\pmb{r}^o})-\lambda P_a+\mu P_d $. From Lagrangian theory, we obtain that in order for a distribution $F_{\pmb{r}^o}$ to be optimal (achieving the maximum), it is necessary and sufficient to
\begin{align}\label{E49}
L^{'}_{F_{\pmb{r}^o}}(F_{\pmb{r}},\lambda,\mu)\leq 0, ~~\forall F_{\pmb{r}}\in \Omega_1\cap\Omega_2.
\end{align}

Following the same approach in \cite{Shamai_BarDavid_1995,Smith:IC:71,AbouFaycal_Trott_Shamai_2001,Tchamkerten_2004,Fahs_AbouFaycal_2012}, it is verified that (\ref{E49}) is equivalent to
\begin{align}\label{E50}
\left\{\begin{array}{ll}
  h(r;F_{\pmb{r}^o})-\lambda r^2+\mu g(r) &=K_{F_{\pmb{r}^o}},~~ r\in \text{supp}\{\pmb{r}^{o}\} \\
  h(r;F_{\pmb{r}^o})-\lambda r^2+\mu g(r) &\leq K_{F_{\pmb{r}^o}},~~ r\in [0,r_p].
\end{array}\right.
\end{align}

Assume that the optimal input $\pmb{r}^o$ contains at least one limit point in its support. This case occurs if support of $\pmb{r}^o$ contains an interval or it is discrete with an infinite number of mass points\footnote{The existence of a limit point in this case follows by Bolzano-Weierstrass theorem.\label{footnote_29}}. Extending the equation in (\ref{E50}) to the complex domain, we have
\begin{align}\label{E51}
  h(z;F_{\pmb{r}^o})=\lambda z^2-\mu g(z)+K_{F_{\pmb{r}^o}},~~z\in\text{Re}(z)>0.
\end{align}
$h(z;F_{\pmb{r}^o})$ is analytic due to analyticity of $K(R,z)$ (see (\ref{E14})) on the domain defined by $\text{Re}(z)>0$. (\ref{E51}) holds if $z$ is the support of $\pmb{r}^o$ on $[0,r_p]$ (due to (\ref{E50})). Hence, by the identity theorem, we have $h(z;F_{\pmb{r}^o})=\lambda z^2-\mu g(z)+ K_{F_{\pmb{r}^o}}$ over the whole domain $\text{Re}(z)>0$ if $z\in \text{supp}\{\pmb{r}^o\}$ is a limit point. In the following, we examine (\ref{E51}) for different range of values for $\lambda\geq 0$ and $\mu\in \mathbb{R}$.
\begin{itemize}
  \item ($\lambda=\mu=0$): Expanding $h(r;F_{\pmb{r}^o})$ from (\ref{E14}), the KKT equality condition in (\ref{E50}) reads as
  \begin{align}\label{E52}
  \int\limits_{0}^{\infty} K(R,r) \ln{\frac{R}{f_{\pmb{R}}(R;F_{\pmb{r}^o})}}dR=H(F_{\pmb{r}}).
  \end{align}
  Noting that the integral transform in (\ref{E52}) is invertible, i.e., the solution is unique (see Appendix \ref{A8}), we have
  \begin{align}
  f_{\pmb{R}}(R;F_{\pmb{r}^o})=Re^{-H(F_{\pmb{r}})},
  \end{align}
  which can be easily verified that is not a legitimate pdf.

  \item ($\lambda>0,\mu=0$): In this case the problem at hand is reduced to the capacity of an AWGN channel under average power and amplitude constraints. In \cite{Shamai_BarDavid_1995}, it is shown that the optimal inputs for this setup are discrete with a finite number of mass points.

  \item ($\lambda\geq 0,\mu>0$): By expanding $h(r;F_{\pmb{r}^o})$ from (\ref{E14}), we have
  \begin{align}\nonumber
  &\int\limits_{0}^{\infty} K(R,r) \ln{\frac{R}{f_{\pmb{R}}(R;F_{\pmb{r}^o})}}dR\\\nonumber
  &=\int\limits_{0}^{\infty} K(R,r) \ln{R}dR\\\label{E95}
  &\quad-\int\limits_{0}^{\infty} K(R,r) \ln{f_{\pmb{R}}(R;F_{\pmb{r}^o})}dR\\\label{E59}
  &>\int\limits_{0}^{1} \ln{R}dR-\int\limits_{0}^{\infty} f_{\pmb{R}}(R;F_{\pmb{r}^o})dR>-2,
  \end{align}
  where the first inequality (\ref{E59}) is due to (\ref{E26}) and $\ln{x}<x$. Therefore, since $h(r;F_{\pmb{r}^o})$ is bounded from below, the optimality conditions in (\ref{E50}) are not met as $r\rightarrow \infty$ (under the assumption of $\mu>0$).

    \item ($\lambda\geq 0,\mu<0$): From Lemma \ref{lem8}, it can be easily verified that $f_{\pmb{R}}(R)$ is in the form of
    \begin{align}\label{E97}
    f_{\pmb{R}}(R;F_{\pmb{r}^o})=R\exp\left\{-\sum\limits_{i=1}^{d}c_iR^{2i}\right\}.
    \end{align}
    According to Lemma \ref{lem9}, (\ref{E97}) is not a legitimate distribution, since it decays faster than $e^{-y^2}$ (recall that $d\geq 2$).
    \end{itemize}

Therefore, the only possibility for the optimal amplitude $\pmb{r}^o$ is to be discrete with a finite number of mass points. We note that, the channel input is indeed continuous due to the uniformly distributed phase.

\qed


\section{Proof of Lemma \ref{Prop1}} \label{app:1}
Considering first the term $\mathbb{E}\mathcal{E}[Y_{\text{rf}}(t)^2]$, we have
\begin{align}\label{eqn:29}
  \mathbb{E}\mathcal{E}&[Y_{\text{rf}}(t)^2] =P+\sigma_n^2.
\end{align}

Considering the term $\mathbb{E}\mathcal{E}[Y_{\text{rf}}(t)^4]$, we have
\begin{align}\label{eqn:5}
\mathbb{E}\mathcal{E}[Y_{\text{rf}}(t)^4] =\frac{3}{2}\mathbb{E}\mathcal{E}\left[|Y(t)|^4\right].
\end{align}
Note that, the signal $|Y(t)|^2$ is real with bandwidth $[-f_w,f_w]$. Hence, it can be represented by its samples taken each $t=1/2f_w$ seconds. Therefore, we have
\begin{align}\label{E77}
|Y(t)|^2=\sum_{k}\pmb{s}_k\text{sinc}(2f_wt-k),
\end{align}
where $\pmb{s}_k \triangleq |Y(k/2f_w)|^2$. Accordingly, (\ref{eqn:5}) reads as
\begin{align}
&\frac{3}{2}\mathbb{E}\mathcal{E}\left[|Y(t)|^4\right]=\lim_{T\rightarrow\infty}\frac{3}{2f_w}\sum_{k}\mathbb{E}[|\pmb{s}_k|^2]\\\label{eqn:6}
&=\lim_{T\rightarrow\infty}\frac{3}{2Tf_w}\sum_{k}\mathbb{E}[|\pmb{s}_{2k+1}|^2]+\frac{3}{2Tf_w}\sum_{k}\mathbb{E}[|\pmb{s}_{2k}|^2]\\\label{E74}
&=\frac{3}{2}\left(\mathbb{E}[|\pmb{s}_{2k+1}|^2]+\mathbb{E}[|\pmb{s}_{2k}|^2]\right).
\end{align}
Note that $\pmb{s}_{2k}= |Y(2k/2f_w)|^2= |\pmb{y}_{k}|^2$. Hence, $\mathbb{E}[|\pmb{s}_{2k}|^2]$ in (\ref{eqn:6}) reads
\begin{align}\label{E85}
\mathbb{E}[|\pmb{s}_{2k}|^2]&=\mathbb{E}[|\pmb{y}_{k}|^4]\\\nonumber
&=\mathbb{E}[|(\pmb{x}+\pmb{n})(\overline{\pmb{x}}+\overline{\pmb{n}})|^2]\\\label{eqn:7}
&=Q+16P+32.
\end{align}

To calculate the term $\mathbb{E}[|\pmb{s}_{2k+1}|^2]$ in (\ref{eqn:6}), we note that the channel's baseband equivalent signal $Y(t)$ can be written as
\begin{align}
Y(t)=\sum_{n}\pmb{x}_n\text{sinc}(f_wt-n)+W(t),
\end{align}
Substituting $t=(2k+1)/f_w$ we have
\begin{align}
\tilde{\pmb{y}}_k&\triangleq Y(t)|_{t=\frac{2k+1}{2f_w}}\\
&=\tilde{\pmb{x}}+\tilde{\pmb{n}}.
\end{align}
where $\tilde{\pmb{x}}\triangleq \sum_{n=-\infty}^{\infty}\pmb{x}_ns_{k-n}$ and $\tilde{\pmb{n}}\triangleq W((2k+1)/2f_w)$. Similarly to (\ref{eqn:7}), we have
\begin{align}
\mathbb{E}[|\pmb{s}_{2k+1}|^2]&=\mathbb{E}[|\tilde{\pmb{y}}_k|^4]\\\label{eqn:8}
&=\tilde{Q}+16\tilde{P}+32,
\end{align}
where $\tilde{Q}=\mathbb{E}[|\tilde{\pmb{x}}|^4],~\tilde{P}=\mathbb{E}[|\tilde{\pmb{x}}|^2]$. For $\tilde{P}$, we have
\begin{align}\label{eqn:30}
\tilde{P}&=\mathbb{E}\bigg[\sum_{n,m}\pmb{x}_n\overline{\pmb{x}_m}s_{k-n}s_{k-m}\bigg]\\\nonumber
&=\sum_{n,m:n=m}\mathbb{E}[|\pmb{x}_n|^2]s_{k-n}^2\\\label{eqn:26}
&\quad+\sum_{n,m:n\neq m}\mathbb{E}[\pmb{x}_n]\mathbb{E}[\overline{\pmb{x}_m}]s_{k-n}s_{k-m}\\
&=S_0P+S_1|\mu|^2\\\label{eqn:28}
&=P,
\end{align}
where in (\ref{eqn:26}) we used the assumption that $\pmb{x}_n$ is i.i.d. with respect to different values of $n$.
For $\tilde{Q}$, we have
\begin{align}\label{eqn:9}
\tilde{Q}&=\mathbb{E}\bigg[\sum_{l,k,d,m}\pmb{x}_l\overline{\pmb{x}_k}\pmb{x}_d\overline{\pmb{x}_m}s_{n-l}s_{n-k}s_{n-d}s_{n-m}\bigg].
\end{align}

Accounting for the different cases for the possible values of $l,k,d,m$, we have
\begin{itemize}
\item If all the indices $l,k,d,m$ are with different values, we have
\begin{align}
\tilde{Q}=|\mu|^4S_2.
\end{align}
\item If $(l=k,~d\neq k,~d=m)$ or $(l=d,~k\neq d,~k=m)$, we have
\begin{align}
\tilde{Q}=P^2S_3.
\end{align}
\item If $(l=m,~k\neq m,~k=d)$, we have
\begin{align}
\tilde{Q}=|P^{'}|^2S_3.
\end{align}
\item If $(l=k,~d\neq m,~d\neq k,~m\neq k)$ or $(l=d,~k\neq m,~k\neq d,~m\neq d)$ or $(k=m,~l\neq d,~l\neq m,~d\neq m)$ or $(d=m,~l\neq k,~l\neq m,~k\neq m)$, we have
\begin{align}
\tilde{Q}=P|\mu|^2S_4.
\end{align}
\item If $(l=m,~k\neq d,~k\neq m,~d\neq m)$, we have
\begin{align}
\tilde{Q}=P^{'}\overline{\mu}^{2}S_4.
\end{align}
\item If $(k=d,~l\neq m,~l\neq d,~m\neq d)$, we have
\begin{align}
\tilde{Q}=\overline{P^{'}}\mu^{2}S_4.
\end{align}
\item If $l=k=d=m$, we have
\begin{align}
\tilde{Q}=QS_5.
\end{align}
\item If $l=k=d\neq m$ or $k=d=m\neq l$, we have
\begin{align}
\tilde{Q}=\overline{T^{'}}\mu S_6.
\end{align}
\item If $l=d=m\neq k$ or $l=k=m\neq d$, we have
\begin{align}
\tilde{Q}=T^{'}\overline{\mu}S_6.
\end{align}
\end{itemize}
In the above expressions we define $P^{'}\triangleq\mathbb{E}[\pmb{x}^2]$, $T^{'}\triangleq\mathbb{E}[|\pmb{x}|^2\pmb{x}]$. Hence, (\ref{eqn:9}) reads
\begin{align}\nonumber
\tilde{Q}&=|\mu|^4S_2+(2P^2+|P^{'}|^2)S_3+(4P|\mu|^2+P^{'}\overline{\mu}^{2}\\\nonumber
&\quad+\overline{P^{'}}\mu^2)S_4+QS_5+2(T^{'}\overline{\mu}+\overline{T^{'}}\mu)S_6\\\nonumber
&=\frac{1}{3}\bigg[Q+4P(P-|\mu|^2)\\\label{eqn:10}
&\quad+2(|P^{'}|^2-\text{Re}\{P^{'}\overline{\mu}^{2}\})+2\text{Re}\{T^{'}\overline{\mu}\}\bigg].
\end{align}

Expanding the terms $|P^{'}|^2-\text{Re}\{P^{'}\overline{\mu}^{2}\}$ and $\text{Re}\{T^{'}\overline{\mu}\}$ in (\ref{eqn:10}), we have
\begin{align}\label{eqn:11}
|P^{'}|^2\!-\!\text{Re}\{P^{'}\overline{\mu}^{2}\}&=(P_r\!-\!P_i)(P_r\!-\!P_i\!-\!(\mu_r^2\!-\!\mu_i^2)),\\\label{eqn:12}
\text{Re}\{T^{'}\overline{\mu}\}&=\mu_r(T_r\!+\!\mu_rP_i)\!+\!\mu_i(T_i\!+\!\mu_iP_r).
\end{align}

Noting that $Q=Q_i+Q_r+2P_rP_i$ and substituting in (\ref{eqn:10}) along with (\ref{eqn:11}) and (\ref{eqn:12}), after some manipulations $\tilde{Q}$ reads
\begin{align}\nonumber
\tilde{Q}&=\frac{1}{3}\big(Q_r+Q_i+2(\mu_rT_r+\mu_iT_i)\\\label{eqn:27}
&\quad+6(P_rP_i+P_r(P_r-\mu_r^2)+P_i(P_i-\mu_i)\big).
\end{align}

Substituting (\ref{eqn:27}), (\ref{eqn:28}) in (\ref{eqn:8}) and substituting the result along with (\ref{eqn:7}) in (\ref{eqn:6}), and adding with (\ref{eqn:29}) yields the result of the Proposition.


\section{Proof of Lemma \ref{Prop2}} \label{app:2}
Note that constraining the input distributions $f_{\pmb{x}}(x)$ to those of non-zero mean Gaussian distributions for each dimension, we have $\text{Re}\{\pmb{x}\}\sim \mathcal{N}(\mu_r,\sigma_r^2)$ and $\text{Im}\{\pmb{x}\}\sim \mathcal{N}(\mu_i,\sigma_i^2)$, where $\sigma_r^2\triangleq P_r-\mu_r^2$ and $\sigma_i^2\triangleq P_i-\mu_i^2$. Therefore, the rate maximization problem reads
\begin{equation}\label{eqn:14}
\begin{aligned}
& \underset{ \mu_r,\mu_i,P_r,P_i}{\text{max}}
& & \frac{f_w}{2}\left(\ln(1+a\sigma_r^2)+\ln(1+a\sigma_i^2)\right) \\
& \text{s.t.}
& &\!\!\!\!\!\!\!\!\!\!\!\!\!\!\!\!\!\! \left\{\begin{array}{l}
      P_r+P_i\leq P_{a}\\
      \alpha (Q+ \tilde{Q})+\beta P+\gamma\geq P_d\\
      \sigma_r^2\geq 0 ,~\sigma_i^2\geq 0
    \end{array}\right.,
\end{aligned}
\end{equation}
where $a\triangleq 2/f_w\sigma_n^2$. Writing the KKT conditions for the optimization problem in (\ref{eqn:14}), we have
\begin{align}\label{eqn:17}
      &\lambda_1(P_r+P_i-P_{a})=0,~ \lambda_1\geq0 \\
      &\lambda_2(\alpha (Q+ \tilde{Q})+\beta P+\gamma-P_d)=0,~ \lambda_2\geq0, \\
      &\zeta_r\sigma_r^2=0,~\zeta_i\sigma_i^2=0,~\zeta_r,\zeta_i\geq 0\\\label{eqn:16}
      &\zeta_r=\frac{-f_wa}{2(1+a\sigma_r^2)}+\lambda_1-\lambda_2 (2\alpha(3P_r+P_i)+\beta),\\\label{eqn:21}
      &\zeta_i=\frac{-f_wa}{2(1+a\sigma_i^2)}+\lambda_1-\lambda_2 (2\alpha(3P_i+P_r)+\beta),\\\label{eqn:15}
      &\frac{f_w a\mu_r}{1+a\sigma_r^2}+8\lambda_2 \alpha\mu_r^3+2\zeta_r\mu_r=0,\\\label{eqn:18}
      &\frac{f_w a\mu_i}{1+a\sigma_i^2}+8\lambda_2 \alpha\mu_i^3+2\zeta_i\mu_i=0,
\end{align}
where in (\ref{eqn:16}) to (\ref{eqn:18}) we used the following
\begin{align}
&\frac{\partial Q}{\partial P_r}=\frac{\partial \tilde{Q}}{\partial P_r}=6P_l+2P_i,\\
&\frac{\partial Q}{\partial P_i}=\frac{\partial \tilde{Q}}{\partial P_i}=6P_i+2P_l,\\
&\frac{\partial Q}{\partial \mu_r}=\frac{\partial \tilde{Q}}{\partial \mu_r}=-8\mu_r^3,\\
&\frac{\partial Q}{\partial \mu_i}=\frac{\partial \tilde{Q}}{\partial \mu_i}=-8\mu_i^3.
\end{align}

It can be easily verified from (\ref{eqn:17}), (\ref{eqn:16}) and (\ref{eqn:21}) that when $\lambda_2=0$, the maximum is achieved when $\mu_r=\mu_i=0$ and $P_r=P_i=\frac{P_{a}}{2}$, yielding $P_{\text{del}}=2\alpha {P_{a}}^2+\beta P_{a} +\gamma$. For positive values of $\lambda_2$ from (\ref{eqn:16}) it is verified that $\lambda_1>0$, which from (\ref{eqn:17}) results that $P_r+P_i=P_{a}$. The condition $P_r+P_i=P_{a}$ reduces the number of variables $P_i,P_r$ to one. Accordingly, since the rate (expansion of the mutual information accounting Gaussian input) is concave wrt $P_i\in[0,P_a]$ attaining its maximum and minimum at $P_i=P_a/2$ and $P_i=0,~P_a$, respectively and  the delivered power $P_{\text{del}}$ is convex wrt $P_i\in[0,P_a]$ attaining its maximum and minimum at $P_i=0,~P_a$ and $P_i=P_a/2$, respectively, the Proposition is proved.


\section{Proof of Lemma \ref{lem6}} \label{A6}
Rewriting the function $I_0(x)$, we have
\begin{align}\label{E33}
  &I_0(x)=\frac{1}{\pi}\int\limits_{0}^{\pi}e^{x\cos(t)}dt\\\label{E32}
  &=\frac{e^x}{\pi\sqrt{x}}\int\limits_{0}^{2\sqrt{x}}\frac{e^{-\frac{u^2}{2}}}{\sqrt{1-\frac{u^2}{4x}}}du\\\label{E34}
  &=\frac{e^x}{\pi\sqrt{x}}\int\limits_{0}^{2\sqrt{ax}}\frac{e^{-\frac{u^2}{2}}}{\sqrt{1-\frac{u^2}{4x}}}du+\frac{e^x}{\pi}\int\limits_{a}^{1}\frac{e^{-2xt}}{\sqrt{t(1-t)}}dt\\\nonumber
  &<\frac{e^x}{\pi\sqrt{x}}\int\limits_{0}^{2\sqrt{ax}}\left(\frac{\hat{a}u}{\sqrt{x}}+1\right)e^{-\frac{u^2}{2}}du\\\label{E30}
  &\quad+\frac{e^x}{\pi}\int\limits_{a}^{1}\frac{e^{-2xt}}{\sqrt{(t-a)(1-t)}}dt,
\end{align}
where (\ref{E33}) is the definition, in (\ref{E32}), we used the transformation $u=2\sqrt{x}\sin(t/2)$, in (\ref{E34}), $0<a<1$ and in the last term of (\ref{E34}), we used the transformation $u^2/2=2xt$. In (\ref{E30}), we used the inequalities $1/\sqrt{1-u^2/4x}<\hat{a}u/\sqrt{x}+1, ~0\leq u\leq 2\sqrt{ax},~ \hat{a}=(1/\sqrt{1-a}-1)/(2\sqrt{a})$\footnote{This can be easily verified by noting that the function $f(u)=\frac{1}{\sqrt{1-u^2/4x}}-(\hat{a}u+1)$ is concave and $f(0)=f(2\sqrt{ax})=0$.\label{footnote_30}}  and $\sqrt{t}>\sqrt{t-a},~t\geq a$, for the first and second terms, respectively. The first integral in (\ref{E30}) is the error function. From \cite[ET I 139(23)]{gradshteyn2007}, the second integral in (\ref{E30}) can be obtained as
\begin{align}\nonumber
\int\limits_{a}^{1}\frac{e^{-2xt}}{\sqrt{(t-a)(1-t)}}dt&=\pi e^{-2x} \Phi(1/2,1;2x(1-a))\\\label{E36}
&<\pi e^{-2ax}.
\end{align}
The inequality in (\ref{E36}) can be easily verified from the definition of $\Phi(\cdot,\cdot;\cdot)$, that is, we have
\begin{align}\label{E35}
\Phi(1/2,1;2x(1-a))\!<\!\Phi(1,1;2x(1-a))\!=\!e^{2x(1-a)},
\end{align}
where the equality in (\ref{E35}) is due to \cite[MO 15]{gradshteyn2007}. Hence, the term in (\ref{E30}) can be further upper bounded by (\ref{E36}) as follows
\begin{align}\label{E39}
  I_0(x)\!<\!e^x\left(\frac{\hat{a}(1-e^{-2ax})}{\pi x}\!+\!\frac{ \mathrm{erf}\left(\sqrt{2ax}\right)}{\sqrt{2\pi x}}\!+\!e^{-2ax}\right).
\end{align}

Since (\ref{E39}) is valid for any $a\in (0,1)$, therefore, the result of the lemma is concluded.

\qed


\section{Proof of Lemma \ref{lem8}} \label{A8}
Using the transform $t=u^2/2$ and \cite[MI 45]{gradshteyn2007}, it is verified that
\begin{align}\nonumber
&\int\limits_{0}^{\infty}R^b K(R,r)dR=\\\label{E56}
&2^{\frac{b}{2}}\Gamma\left(\frac{b}{2}+1\right)e^{-\frac{r^2}{2}}\Phi\left(\frac{b}{2}+1,1;\frac{r^2}{2}\right),
\end{align}
for $0\leq r<\infty,~b>-2$. By substituting $G(R)$ in (\ref{E55}), and using (\ref{E56}), we have
\begin{align}\nonumber
&\sum\limits_{i=0}^{n}c_i\int\limits_{0}^{\infty}R^{2i} K(R,r)dR\\\label{E62}
&\quad=\sum\limits_{i=0}^{n}c_i2^{i}i!e^{-\frac{r^2}{2}}\Phi\left(i+1,1;\frac{r^2}{2}\right).
\end{align}
The function $\Phi\left(i+1,1;\frac{r^2}{2}\right)$ can be found for integer values of $i$ using the following two properties of \textit{Confluent Hypergeometric functions} (see \cite[MO 15, MO 112]{gradshteyn2007})
\begin{subequations}\label{E60}
\begin{align}
  \Phi\left(i,i;x\right) &= e^x,~i=1,2,\ldots, \\\label{E60b}
  \Phi\left(a+1,b;x\right) &=\frac{x}{b}\Phi\left(a+1,b+1;x\right)+\Phi\left(a,b;x\right).
\end{align}
\end{subequations}

Denoting $e^{-\frac{r^2}{2}}\Phi\left(i,k;\frac{r^2}{2}\right)\triangleq \Phi_{r}(i,k)$ for $i=1,2,\ldots$ and $k=0,\ldotp,i-1$, we have
\begin{subequations}
\begin{align}
\Phi_{r}(i,i)&\!=\!1,\\
\Phi_{r}(i,i\!-\!1)&\!=\!\frac{r^2}{2(i\!-\!1)}\Phi_{r}(i,i)\!+\!\Phi_{r}(i\!-\!1,i\!-\!1)\\
&\!=\!\frac{r^2}{2(i\!-\!1)}\!+\!1,\\
\Phi_{r}(i,i\!-\!2)&\!=\!\frac{r^2}{2(i\!-\!2)}\Phi_{r}(i,i\!-\!1)+\Phi_{r}(i\!-\!1,i\!-\!2)\\
&\!=\!\frac{r^2}{2(i\!-\!2)}\left(\frac{r^2}{2(i\!-\!1)}\!+\!1\right)\!+\!\frac{r^2}{2(i\!-\!2)}\!+\!1,\\\nonumber
\vdots\\\label{E73a}
\Phi_{r}(i,k)&\!=\!\frac{r^2}{2k}\Phi_r(i,k\!+\!1)\!+\!\Phi_r(i\!-\!1,k),\\\nonumber
\vdots\\
\Phi_{r}(i,2)&\!=\!\frac{r^2}{4}\Phi_r(i,3)\!+\!\Phi_r(i\!-\!1,2),\\
\Phi_{r}(i,1)&\!=\!\frac{r^2}{2}\Phi_r(i,2)\!+\!\Phi_r(i\!-\!1,1).
\end{align}
\end{subequations}
Note that for example in (\ref{E73a}), both $\Phi_r(i,k+1),~\Phi_r(i-1,k)$ can be obtained from the previous stage. Also, it is verified that $\Phi_r(i,k)$ is a polynomial of degree $2(i-k),~1\leq k\leq i$, i.e., the degree of the polynomial depends on the difference of the arguments $i,~k$. Therefore, $\Phi_{r}(i,1)$ is a polynomial of degree $2(i-1)$.

Using the aforementioned approach, in the following, we have calculated $\Phi_{r}(i,1)$ for $i=2,\ldots,6$
\begin{subequations}\label{E61}
\begin{align}
  \Phi_{r}(2,1)&= \frac{r^2}{2}+1, \\
  \Phi_{r}(3,1)&=\frac{r^4}{8}+r^2+1,\\
  \Phi_{r}(4,1)&=\frac{r^6}{48}+\frac{3r^4}{8}+\frac{3r^2}{2}+1,\\
  \Phi_{r}(5,1)&=\frac{r^8}{384}+\frac{r^6}{12}+\frac{3r^4}{4}+2r^2+1,\\
  \Phi_{r}(6,1)&=\frac{r^{10}}{3840}+\frac{5r^8}{384}+\frac{5r^6}{24}+\frac{5r^4}{4}+\frac{5^2}{2}+1.
\end{align}
\end{subequations}
Therefore, $c_i$s can be simply found by comparing the RHS of (\ref{E62}) with $g(r)$. Uniqueness of the coefficients $c_i$ is guaranteed by the fact that the integral transform in (\ref{E55}) is invertible. To verify the invertibility of (\ref{E55}), consider the following transform
\begin{align}\label{E43}
V(r)=\int\limits_{0}^{\infty}K(R,r)S(R)dR,
\end{align}
where $S(R)$ is restricted to be a polynomial with a finite degree in order to guarantee the existence of the transform. It is enough to show that $S(R)=0$ if and only if $V(r)=0$\footnote{Note that an integral transform functional $\mathcal{I}(f)$ is invertible when $\mathcal{I}(f_1-f_2)=0\overset{\textrm{iff}}\leftrightarrow f_1=f_2$.\label{footnote_31}}. It is easily verified that $S(R)=0$ yields $V(r)=0$. For the converse, assume $V(r)=0$. By taking the second integral over $r$ as below, we have
\begin{align}\label{E41}
\int\limits_{0}^{\infty}re^{-sr^2}\int\limits_{0}^{\infty}K(R,r)S(R)dRdr=0,~~s\geq0.
\end{align}
By changing the order of the integrals in (\ref{E41}) (This is validated by our assumption on $S(R)$ and due to Fubini's theorem), we have
\begin{align}\nonumber
&\int\limits_{0}^{\infty}\int\limits_{0}^{\infty}re^{-sr^2}K(R,r)S(R)drdR\\\label{E44}
&=\int\limits_{0}^{\infty}\frac{Re^{-\frac{R^2}{2}}S(R)}{2}\int\limits_{0}^{\infty}e^{-t\left(s+\frac{1}{2}\right)}I_0\left(R\sqrt{t}\right)dtdR\\\label{E45}
&=\frac{1}{1+2s}\int\limits_{0}^{\infty}Re^{-R^2\left(\frac{1+s}{1+2s}\right)}S(R)dR=0,~~s\geq 0,
\end{align}
where (\ref{E44}) is obtained by expanding $K(R,r)$ and transformation $r^2=t$. (\ref{E45}) is obtained using \cite[ET I 197(20)a, MO 115, MO 15]{gradshteyn2007}. From (\ref{E45}) it is verified that (\ref{E41}) is valid only if $S(R)=0$.

\qed


\section{Proof of Lemma \ref{lem4}} \label{A4}
Solving $\frac{\partial{K(R,r)}}{\partial{r}}=0$, we have
\begin{align}\label{E6}
r^{*}=R\frac{I_0^{'}(r^{*}R)}{I_0(r^{*}R)}=R\frac{I_1(r^{*}R)}{I_0(r^{*}R)},
\end{align}
where the second equality in (\ref{E6}) is due to the equality $I_0^{'}(x)=I_1(x)$. Using the inequalities $\frac{I_0^{'}(x)}{I_0(x)}<1$ and $\frac{I_1(x)}{I_0(x)}\geq\frac{x}{x+1}$ from \cite{Laforgia_2010}, we have
\begin{align}\label{E10}
R-\frac{1}{R}\leq r^{*}<R.
\end{align}

Note that for $R>\sqrt{2}$ we have $r^{*}>R/2$. Rewriting $f_{\pmb{R}}(R,F_{\pmb{r},n})$ for $R>2$, we have
\begin{align}
&f_{\pmb{R}}(R,F_{\pmb{r},n})=\int\limits_{0}^{\infty}K(R,r)dF_{\pmb{r}}(r)\\
&=\int\limits_{0}^{\frac{R}{2}}K(R,r)dF_{\pmb{r}}(r)+\int\limits_{\frac{R}{2}}^{\infty}K(R,r)dF_{\pmb{r}}(r)\\
&<K\left(R,R/2\right)\textrm{Pr}\left(\pmb{r}\!\leq\!\frac{R}{2}\right)\!+\!K\left(R,r^{*}\right) \textrm{Pr}\left(\pmb{r}\!>\!\frac{R}{2}\right)\\\label{E27}
&<K\left(R,R/2\right)+\frac{4P_a}{R^2}\\
&=Re^{-\frac{5R^2}{8}}I_0(R^2/2)+\frac{4P_a}{R^2}\\\label{E28}
&<Re^{-\frac{R^2}{8}}+\frac{4P_a}{R^2}\\\label{E29}
&<\frac{128R}{R^4}+\frac{4P_a}{R^2}\\
&<\frac{128}{R^2}+\frac{4P_a}{R^2}\\
&=\frac{128+4 P_a}{R^2},
\end{align}
where (\ref{E27}) is due to the Markov's inequality and (\ref{E26}). (\ref{E28}) is due to $I_0(x)<e^x$. (\ref{E29}) is due to $e^{-x}\leq \frac{k!}{x^k}$ for any nonnegative integer $k$ (here $k=2$).

Finally, from (\ref{E31}) and the inequality $|x\ln{x}|<4x^{\frac{3}{4}},~0\leq x \leq 1$ we have
\begin{align}
&|f_{\pmb{R}}(R,F_{\pmb{r},n})\ln f_{\pmb{R}}(R,F_{\pmb{r},n})|< 4f_{\pmb{R}}(R,F_{\pmb{r},n})^{\frac{3}{4}}\\
&\quad\quad< g(R)=\left\{\begin{array}{cc}
     4 & R\leq 2 \\
     \frac{c}{R^{\frac{3}{2}}} & R>2
   \end{array}\right.,
\end{align}
where $c=4(128+4P_a)^{\frac{3}{4}}$. It is easy to verify that $g(R)$ is integrable.

\qed

\section{Proof of Lemma \ref{lem5}} \label{A5}
\textit{1) Existence:} Rewriting $|H(F_{\pmb{r}})|$ in (\ref{E8}), we have
\begin{align}\nonumber
 | H(F_{\pmb{r}})|\leq& \int\limits_{0}^{\infty} f_{\pmb{R}}(R;F_{\pmb{r}})\ln \frac{1}{f_{\pmb{R}}(R;F_{\pmb{r}})} dR\\\label{E13}
 &+ \int\limits_{0}^{\infty} f_{\pmb{R}}(R;F_{\pmb{r}})|\ln R |dR.
\end{align}
The first term in the RHS of (\ref{E13}) is the entropy of the random variable $\pmb{R}$, which exists and is finite due to Lemma \ref{lem4} and is always positive due to (\ref{E31}). For the second term in the RHS of (\ref{E13}) and for any $F_{\pmb{r}}\in \Omega_1\cap\Omega_2$ we have
\begin{align}\nonumber
\int\limits_{0}^{\infty} f_{\pmb{R}}(R;F_{\pmb{r}})|\ln R |dR&=\int\limits_{0}^{1} f_{\pmb{R}}(R;F_{\pmb{r}})|\ln R |dR\\\label{E25}
&\quad+\int\limits_{1}^{\infty} f_{\pmb{R}}(R;F_{\pmb{r}})\ln RdR.
\end{align}
The first term in (\ref{E25}) is bounded by noting that $\int_{0}^{1} f_{\pmb{R}}(R;F_{\pmb{r}})\ln R dR<0$ and due to
\begin{align}\label{E40}
\int\limits_{0}^{1} f_{\pmb{R}}(R;F_{\pmb{r}})\ln R dR&>\int\limits_{0}^{1} \ln R dR=-1,
\end{align}
where the inequality in (\ref{E40}) is due to (\ref{E31}). The second term in (\ref{E25}) is bounded due to the inequality $\ln{x}<\sqrt{x}$ and the following lemma
\begin{lem}\label{lem3}
The expectation $\mathbb{E}[\pmb{R}^\alpha]$, $0\leq \alpha<1$ for any $F_{\pmb{r}}\in \Omega_1\cap\Omega_2$ exists and is bounded.
\end{lem}
\textit{Proof}: See Appendix \ref{A3}.

Existence of (\ref{E25}) validates existence of $H(F_{\pmb{r}})$ and this concludes the proof.

\textit{2) Continuity:} Let $F_{\pmb{r},n}\overset {w}\rightarrow F$. Using the weak topology, the continuity of $H(F_{\pmb{r}})$ is equivalent to
\begin{align}
F_{\pmb{r},n}\overset {w}\rightarrow F \Longrightarrow H(F_{\pmb{r},n})\rightarrow H(F_{\pmb{r}}).
\end{align}
Therefore, we have
\begin{align}\nonumber
&\lim\limits_{n}H(F_{\pmb{r},n})\\\label{E15}
&=-\lim\limits_{n}\int\limits_{0}^{\infty} f_{\pmb{R}}(R,F_{\pmb{r},n})\ln \frac{f_{\pmb{R}}(R,F_{\pmb{r},n})}{R} dR\\\label{E16}
&=-\int\limits_{0}^{\infty} \lim\limits_{n} f_{\pmb{R}}(R,F_{\pmb{r},n})\ln \frac{f_{\pmb{R}}(R,F_{\pmb{r},n})}{R} dR\\\label{E17}
&=\int\limits_{0}^{\infty} f_{\pmb{R}}(R;F_{\pmb{r}})\ln \frac{f_{\pmb{R}}(R;F_{\pmb{r}})}{R} dR\\\label{E18}
&=H(F_{\pmb{r}}),
\end{align}
where (\ref{E15}) and (\ref{E18}) are definitions. (\ref{E16}) is due to Lebesgue Dominated Convergence Theorem and absolute integrability of the integrand in (\ref{E15}) due to Lemma \ref{lem4}. (\ref{E17}) is due to continuity of $x\ln x$.

\textit{3) Strict concavity:} Concavity follows by noting that in (\ref{E13}), the first term is the entropy function and therefore concave with respect to the distribution function $f_{\pmb{R}}(R;F_{\pmb{r}})$, and the second term is a linear function of $f_{\pmb{R}}(R;F_{\pmb{r}})$. Strict concavity follows by noting that the transform
\begin{align}
f_{\pmb{R}}(R;F_{\pmb{r}})=\int\limits_{0}^{\infty}K(R,r)dF_{\pmb{r}}(r),
\end{align}
is invertible (for the proof see \cite[Appendix II]{Shamai_BarDavid_1995}).

\textit{3) Weak differentiability:} The proof for weak differentiability is the same as \cite[Proposition 4]{Shamai_BarDavid_1995} or \cite[AppendixII.B]{AbouFaycal_Trott_Shamai_2001}. For brevity we avoid the details. It can be verified that applying weak derivative over (\ref{E13}) yields
\begin{align}\nonumber
&H^{'}_{F_{\pmb{r}^o}}(F_{\pmb{r}}) \\
&=\lim\limits_{\theta \rightarrow0}\frac{H((1-\theta)F^{0}_{\pmb{r}}+\theta F_{\pmb{r}})-H(F^{0}_{\pmb{r}}) }{\theta}, ~\theta\in[0,1]\\
&=\int\limits_{0}^{\infty}h(r;F_{\pmb{r}^o})dF_{\pmb{r}}-H(F^{0}_{\pmb{r}}),
\end{align}
where $h(r;F_{\pmb{r}^o})$ is defined as in (\ref{E14}).

We conclude the proof by noting that the integral transform in (\ref{E14}) is invertible (see Appendix \ref{A8}).


\section{Proof of Lemma \ref{lem3}} \label{A3}
For $\mathbb{E}[\pmb{R}^\alpha]$ we have
\begin{align}
\mathbb{E}[\pmb{R}^\alpha]=&\int\limits_{0}^{\infty}R^\alpha f_{\pmb{R}}(R;F_{\pmb{r}})dR\\
=&\int\limits_{0}^{\infty}\int\limits_{0}^{\infty}R^\alpha  K(R,r)dF_{\pmb{r}}(r)dR\\\nonumber
=&\int\limits_{0}^{2}\int\limits_{0}^{\infty}R^\alpha  K(R,r)dF_{\pmb{r}}(r)dR\\\nonumber
&+\int\limits_{2}^{\infty}\int\limits_{0}^{1}R^\alpha  K(R,r)dF_{\pmb{r}}(r)dR\\\label{E21}
&+\int\limits_{2}^{\infty}\int\limits_{1}^{\infty}R^\alpha  K(R,r)dF_{\pmb{r}}(r)dR,
\end{align}
where we have divided the integrals due to the similar reason explained in Appendix (\ref{A4}) (see equation (\ref{E10})).

For the first integral in the RHS of (\ref{E21}) we have
\begin{align}\nonumber
&\int\limits_{0}^{2}\int\limits_{0}^{\infty}R^\alpha  K(R,r)dF_{\pmb{r}}(r)dR\\\label{E22}
&<\int\limits_{0}^{2}R^\alpha dR=\frac{2^{1+\alpha}}{1+\alpha}<\infty, ~\alpha\geq 0,
\end{align}
where the inequality is due to (\ref{E26}). For the second integral in the RHS of (\ref{E21}) we have
\begin{align}\nonumber
&\int\limits_{2}^{\infty}\int\limits_{0}^{1}R^\alpha  K(R,r)dF_{\pmb{r}}(r)dR\\\label{E9}
&<\int\limits_{2}^{\infty}\int\limits_{0}^{1}R^\alpha K(R,1)dF_{\pmb{r}}(r)dR\\\label{E23}
&<\int\limits_{0}^{\infty}R^{1+\alpha} e^{-\frac{(R-1)^2}{2}}<\infty, ~\alpha\geq 0,
\end{align}
where in (\ref{E9}) we used $K(R,r)\leq K(R,1)$ for $R\geq 2,r\leq1$ due to (\ref{E10}). In (\ref{E23}) we used the inequality $I_0(x)<e^{x}$.

Note that for $R\geq2$, it is easy to verify that $1+R/4\leq R-1/R$. This along with (\ref{E10}), guarantee that
\begin{align}\label{E11}
K(R,r)\leq K(R,1+R/4),~~R\geq 2,~r\leq 1+R/4.
\end{align}
Therefore, for the third integral in the RHS of (\ref{E21}) we have
\begin{align}
&\int\limits_{2}^{\infty}\int\limits_{1}^{\infty}R^\alpha  K(R,r)dF_{\pmb{r}}(r)dR\\
&<\int\limits_{2}^{\infty}\int\limits_{1}^{1+\frac{R}{4}}R^\alpha  K(R,r)dF_{\pmb{r}}(r)dR\\
&\quad+\int\limits_{2}^{\infty}\int\limits_{1+\frac{R}{4}}^{\infty}R^\alpha  K(R,r) dF_{\pmb{r}}(r)dR\\\label{E12}
&<\int\limits_{2}^{\infty}R^\alpha  K(R,1+R/4) dR\\
&\quad +\int\limits_{2}^{\infty}R^\alpha \textrm{Pr}(r>1+R/4)dR\\\label{E24}
&<\int\limits_{2}^{\infty}\frac{R^{\alpha+\frac{1}{2}}}{1+\frac{R}{4}} e^{-\frac{(3R-4)^2}{32}} dR\\
&\quad+\int\limits_{2}^{\infty}\frac{P_a R^\alpha}{(1+R/4)^2}dR<\infty,~0\leq \alpha <1,
\end{align}
where (\ref{E12}) is due to (\ref{E11}) and (\ref{E26}). In (\ref{E24}) we used Markov's inequality. From (\ref{E22}), (\ref{E23}) and (\ref{E24}), it can be easily verified that $E[\pmb{R}^\alpha]$ for $0\leq \alpha <1$ exists, which also concludes the result of Lemma \ref{lem3}.

\qed


\section{Upperbound for $K(R,r)$} \label{App_bounding}
Using the upperbound in (\ref{E37}) with $a=0.5$ (accordingly $\hat{a}=0.293$), $K(R,r)$ is upperbounded as
\begin{align}\nonumber
K(R,r)\leq&  R e^{-\frac{R^2+r^2}{2}}+\frac{\mathrm{erf}(\sqrt{rR})R e^{-\frac{R^2+r^2}{2}} }{\sqrt{2\pi rR}}\\\label{K_1}
&+ \frac{0.293(e^{rR}-1)e^{-\frac{R^2+r^2}{2}}}{\pi r} .
\end{align}
In the following, we bound each term in (\ref{K_1}), separately.
\begin{itemize}
\item First term on the RHS of the inequality (\ref{K_1}): Differentiating with respect to $R$, it is verified that the maximum occurs at $R=1$. We get
\begin{align}\label{K_2}
R e^{-\frac{R^2+r^2}{2}} \leq e^{-\frac{1+r^2}{2}}\leq e^{-\frac{1}{2}} < 0.6066.
\end{align}

\item Second term on the RHS of the inequality (\ref{K_1}): We first note that (after taking the derivative and some manipulations) the function $\frac{\mathrm{erf}(x)}{x}$ is decreasing having its supremum at $x=0$. Accordingly, the function $\frac{\mathrm{erf}(\sqrt{x})}{\sqrt{x}}$ is decreasing\footnote{Note that if $f(x)$ is decreasing and $g(x)$ is increasing with respect to $x$, then, f(g(x)) is decreasing.\label{footnote_32}}, and having its supremum at $x=0$ as well\footnote{Note that $\lim_{x\rightarrow 0 }\frac{\mathrm{erf}(\sqrt{x})}{\sqrt{x}}= \frac{2}{\sqrt{\pi}}$\label{footnote_33}}. Therefore, we get
    \begin{align}\label{K_9}
    \frac{\mathrm{erf}(\sqrt{rR})R e^{-\frac{R^2+r^2}{2}} }{\sqrt{2\pi rR}}< \frac{\sqrt{2}\cdot 0.6066}{\pi }<0.274,
    \end{align}
where for the inequality (\ref{K_9}), we also used (\ref{K_2}).

\item Third term on the RHS of the inequality (\ref{K_1}): First assume the range $rR\leq 1$. We get
\begin{align}\nonumber
&\frac{0.293}{\pi}\cdot\frac{e^{rR}-1}{ rR}\cdot Re^{-\frac{R^2+r^2}{2}} \\\nonumber
&\leq \frac{0.293}{\pi}\cdot(e-1)\cdot 0.6066\\\label{K_3}
& < 0.0973,
\end{align}
where the inequality in (\ref{K_3}) is due to (\ref{K_2}) and noting that the function $\frac{e^{rR}-1}{ rR}$ is increasing with respect to $rR\leq 1$ (and its maximum occurs at $rR=1$).

For the range $rR\geq 1$ and $r\geq 1$, we get
\begin{align}\label{tag2}
\frac{0.293(e^{rR}-1)e^{-\frac{R^2+r^2}{2}}}{\pi r} &\leq \frac{0.293}{\pi}\cdot\frac{e^{rR}e^{-\frac{R^2+r^2}{2}}}{ r} \\\label{tag3}
&= \frac{0.293}{\pi}\cdot\frac{e^{-\frac{(r-R)^2}{2}}}{ r}\\\label{K_7}
&\leq \frac{0.293}{\pi}< 0.0955.
\end{align}
And finally, for the range $rR\geq 1$ and $r\leq 1$ (therefore $R\geq 1$), we get
\begin{align}\nonumber
&\frac{0.293}{\pi}\cdot\frac{e^{rR}-1}{ r}\cdot e^{-\frac{R^2+r^2}{2}} \\\label{K_4} &=\frac{0.293}{\pi}\cdot\left(\sum\limits_{i=1}^{\infty}\frac{r^{i-1}R^i}{i!}\right)\cdot e^{-\frac{R^2}{2}}\cdot e^{-\frac{r^2}{2}}\\\label{tag4}
&=\frac{0.293}{\pi}\cdot\left(\sum\limits_{i=1}^{\infty}\frac{r^{i-1}e^{-\frac{r^2}{2}}R^i}{i!}\right)\cdot e^{-\frac{R^2}{2}}\\\label{K_5}
&\leq \frac{0.293}{\pi}\cdot\left(R +0.62\sum\limits_{i=2}^{\infty}\frac{R^i}{i!}\right)\cdot e^{-\frac{R^2}{2}}\\\label{tag5}
&=\frac{0.293}{\pi}\cdot\left(\!Re^{-\frac{R^2}{2}} \!+\!0.62e^{-\frac{R^2}{2}}\cdot\left(e^R\!-\!R\!-\!1\right)\!\right)\\\label{K_6}
&\leq \frac{0.293}{\pi}\cdot\left(0.62 +0.62\cdot0.7\right)< 0.099,
\end{align}
where in (\ref{K_4}), we used Taylor expansion of the function $\frac{e^{rR}-1}{ r}$. (\ref{K_5}) is due to $r^{i-1}e^{-\frac{r^2}{2}}\leq1$ for $i=1$ and $r^{i-1}e^{-\frac{r^2}{2}}<0.62$ for $i\geq 2$ (in the range $r\leq 1$). (\ref{K_6}) is due to $Re^{-\frac{R^2}{2}}<0.62$ and $e^{-\frac{R^2}{2}}(e^R-R-1) < 0.7$.

According to (\ref{K_3}), (\ref{K_7}) and (\ref{K_6}), the third term in the RHS of (\ref{K_1}) is upperbounded as
\begin{align}\nonumber
\frac{0.293}{\pi}&\cdot\frac{e^{rR}-1}{ rR}\cdot Re^{-\frac{R^2+r^2}{2}} \\\label{K_10}
&< \max\{0.0973,0.0955,0.099\}=0.099.
\end{align}
\end{itemize}
According to (\ref{K_2}), (\ref{K_9}) and (\ref{K_10}) we get
\begin{align}\label{tag6}
K(R,r)< 0.6066+0.274 +0.099<1 .
\end{align}
\qed


\section{Proof of Lemma \ref{Lemma1}} \label{app:3}
In the following, if not mentioned, the summations are from $-\infty$ to $\infty$. We have
\begin{align}
T_0&=\sum_l a_l\\
&=\frac{1}{\pi}\sum_{l}\frac{(-1)^l}{\left(\frac{1}{2}+l\right)}\\
&= \frac{2}{\pi}\bigg[\sum_{l=-\infty}^{-1}\frac{(-1)^l}{(2l+1)}+\sum_{0}^{\infty}\frac{(-1)^l}{(2l+1)}\bigg]\\\label{eqn:19}
&=\frac{2}{\pi}\bigg(\frac{\pi}{4}+\frac{\pi}{4}\bigg)=1,\\
S_0&=\sum_l a_l^2\\
&=\sum_{l}\frac{(-1)^{2l}}{\pi^2 \left(\frac{1}{2}+l\right)^2}\\
&=\frac{4}{\pi^2}\sum_l\frac{1}{(2l+1)^2}\\
&= \frac{4}{\pi^2}\bigg[\sum_{l=-\infty}^{-1}\frac{1}{(2l+1)^2}+\sum_{0}^{\infty}\frac{1}{(2l+1)^2}\bigg]\\
&=\frac{4}{\pi^2}\bigg(\frac{\pi^2}{8}+\frac{\pi^2}{8}\bigg)=1,\\
T_1&=\sum_l a_l^3\\
&=\sum_{l}\frac{(-1)^{3l}}{\pi^3 \left(\frac{1}{2}+l\right)^3}\\
&=\frac{8}{\pi^3}\bigg[\sum_{l=-\infty}^{-1}\frac{1}{(2l+1)^3}+\sum_{0}^{\infty}\frac{1}{(2l+1)^3}\bigg]\\
&=\frac{8}{\pi^3}\bigg(\frac{\pi^3}{32}+\frac{\pi^3}{32}\bigg)=\frac{1}{2},\\
S_5&=\sum_l a_l^4\\
&=\sum_{l}\frac{(-1)^{4l}}{\pi^4 \left(\frac{1}{2}+l\right)^4}\\
&=\frac{16}{\pi^4}\sum_l\frac{1}{(2l+1)^4}\\
&= \frac{16}{\pi^4}\bigg[\sum_{l=-\infty}^{-1}\frac{1}{(2l+1)^4}+\sum_{0}^{\infty}\frac{1}{(2l+1)^4}\bigg]\\
&=\frac{16}{\pi^4}\bigg(\frac{\pi^4}{96}+\frac{\pi^4}{96}\bigg)=\frac{1}{3},\\
S_1&=\sum_{l} \sum_{k,k\neq l}a_la_k\\
&=\sum_{l} a_l\bigg(\sum_{k}a_k-a_l\bigg)\\
&=\left(\sum_l a_l\right)^2-\sum_l a_l^2\\
&=1-1=0,\\
S_3&=\sum_{l} \sum_{k,k\neq l}a_l^2a_k^2\\
&=\sum_{l} a_l^2\bigg(\sum_{k}a_k^2-a_l^2\bigg)\\
&=\left(\sum_l a_l^2\right)^2-\sum_l a_l^4\\
&=1-\frac{1}{3}=\frac{2}{3},\\
S_6&=\sum_{l} \sum_{k,k\neq l}a_l^3a_k\\
&=\sum_{l} a_l^3\bigg(\sum_{k}a_k-a_l\bigg)\\
&=\frac{1}{2}-\frac{1}{3}=\frac{1}{6},\\
S_4&=\sum_{l}\sum_{k,k\neq l}\sum_{\substack{d,d\neq l\\d\neq k}}a_l^2a_ka_d\\
&=\sum_{l}\sum_{k,k\neq l} a_l^2 a_k \bigg(\sum_d a_d-a_l-a_k\bigg)\\
&=\sum_{l} a_l^2\bigg((1-a_l)\sum_{k,k\neq l}a_k-\sum_{k,k\neq l}a_k^2\bigg)\\
&=\sum_{l} a_l^2\bigg((1-a_l)^2-(1-a_l^2)\bigg)\\
&=\sum_l 2a_l^2(a_l^2-a_l)\\
&=2\left(\frac{1}{3}-\frac{1}{2}\right)=-\frac{1}{3},\\
S_2&=\sum_{l}\sum_{k,k\neq l}\sum_{\substack{d,d\neq l\\d\neq k}}\sum_{\substack{m,m\neq d\\m\neq l\\m\neq k}}a_la_ka_da_m\\
&=\sum_{l}\sum_{k,k\neq l}\sum_{\substack{d,d\neq l\\d\neq k}}a_la_ka_d(1-a_d-a_l-a_k)\\
&=\sum_{l}\sum_{k,k\neq l}a_la_k\bigg( (1\!-\!a_l\!-\!a_k)\sum_{\substack{d,d\neq l\\d\neq k}}a_d\!-\!\sum_{\substack{d,d\neq l\\d\neq k}}a_d^2\bigg)\\
&=\sum_{l}\sum_{k,k\neq l}a_la_k\bigg( (1\!-\!a_l\!-\!a_k)^2\!-\!(1\!-\!a_l^2\!-\!a_k^2)\bigg)\\\nonumber
&=\sum_{l}a_l\bigg( 2a_l(a_l-1)(1-a_l)\\
&\quad+\sum_{k,k\neq l}2a_k(a_k^2+a_la_k-a_k)\bigg)\\
&=\sum_{l}a_l(-6a_l^3+6a_l^2-1))\\\label{E99}
&=-\frac{6}{3}+\frac{6}{2}-1=0.
\end{align}

\bibliographystyle{ieeetran}
\bibliography{ref}

\begin{thebibliography}{10}
\providecommand{\url}[1]{#1}
\csname url@samestyle\endcsname
\providecommand{\newblock}{\relax}
\providecommand{\bibinfo}[2]{#2}
\providecommand{\BIBentrySTDinterwordspacing}{\spaceskip=0pt\relax}
\providecommand{\BIBentryALTinterwordstretchfactor}{4}
\providecommand{\BIBentryALTinterwordspacing}{\spaceskip=\fontdimen2\font plus
\BIBentryALTinterwordstretchfactor\fontdimen3\font minus
  \fontdimen4\font\relax}
\providecommand{\BIBforeignlanguage}[2]{{%
\expandafter\ifx\csname l@#1\endcsname\relax
\typeout{** WARNING: IEEEtran.bst: No hyphenation pattern has been}%
\typeout{** loaded for the language `#1'. Using the pattern for}%
\typeout{** the default language instead.}%
\else
\language=\csname l@#1\endcsname
\fi
#2}}
\providecommand{\BIBdecl}{\relax}
\BIBdecl

\bibitem{Varasteh_Rassouli_Clerckx_ITW_2017}
M.~Varasteh, B.~Rassouli, and B.~Clerckx, ``Wireless information and power
  transfer over an awgn channel: Nonlinearity and asymmetric gaussian
  signaling,'' in \emph{IEEE Inf. Theory Workshop (ITW)}, Nov. 2017, pp.
  181--185.

\bibitem{Clerckx_Zhang_Schober_Poor}
B.~Clerckx, R.~Zhang, R.~Schober, D.~W.~K. Ng, D.~I. Kim, and H.~V. Poor,
  ``Fundamentals of wireless information and power transfer: From rf energy
  harvester models to signal and system designs,'' \emph{IEEE Journal on
  Selected Areas in Communications}, vol.~37, no.~1, pp. 4--33, Jan. 2019.

\bibitem{Trotter_Griffin_Durgin_2009}
M.~S. Trotter, J.~D. Griffin, and G.~D. Durgin, ``Power-optimized waveforms for
  improving the range and reliability of rfid systems,'' in \emph{IEEE Int.
  Conf. RFID}, Apr. 2009, pp. 80--87.

\bibitem{Clerckx_Bayguzina_2016}
B.~Clerckx and E.~Bayguzina, ``Waveform design for wireless power transfer,''
  \emph{IEEE Trans. Signal Processing}, vol.~64, no.~23, pp. 6313--6328, Dec.
  2016.

\bibitem{Clerckx_Bayguzina_2017}
------, ``Low-complexity adaptive multisine waveform design for wireless power
  transfer,'' \emph{IEEE Antennas and Wireless Propagation Letters}, vol.~16,
  pp. 2207--2210, 2017.

\bibitem{Kim_Clerckx_Mitcheson_arxive}
J.~Kim, B.~Clerckx, and P.~D. Mitcheson, ``Prototyping and experimentation of a
  closed-loop wireless power transmission with channel acquisition and waveform
  optimization,'' in \emph{IEEE Wireless Power Transfer Conference (WPTC)}, May
  2017, pp. 1--4.

\bibitem{Zhang_Keong_2013}
R.~Zhang and C.~Keong, ``Mimo broadcasting for simultaneous wireless
  information and power transfer,'' \emph{IEEE Trans. Wireless Commun.},
  vol.~12, no.~5, pp. 1989--2001, May 2013.

\bibitem{Zeng_Clerckx_Zhang2017}
Y.~Zeng, B.~Clerckx, and R.~Zhang, ``Communications and signals design for
  wireless power transmission,'' \emph{IEEE Trans. Commun.}, vol.~65, no.~5,
  pp. 2264--2290, May 2017.

\bibitem{Grover_Sahai_2010}
P.~Grover and A.~Sahai, ``Shannon meets tesla: Wireless information and power
  transfer,'' in \emph{IEEE Int. Sym. Inf. Theory}, Jun. 2010, pp. 2363--2367.

\bibitem{Park_Clerckx_2013}
J.~Park and B.~Clerckx, ``Joint wireless information and energy transfer in a
  two-user mimo interference channel,'' \emph{IEEE Trans. Wireless Commun.},
  vol.~12, no.~8, pp. 4210--4221, Aug. 2013.

\bibitem{Clerckx_2016}
B.~Clerckx, ``Wireless information and power transfer: Nonlinearity, waveform
  design, and rate-energy tradeoff,'' \emph{IEEE Trans. Signal Processing},
  vol.~66, no.~4, pp. 847--862, Feb. 2018.

\bibitem{Shannon_1948}
C.~E. Shannon, ``A mathematical theory of communication,'' \emph{Bell Syst.
  Tech. J.}, vol.~27, pp. 379--423 and 623--656, Jul. and Oct. 1948.

\bibitem{Smith:IC:71}
J.~G. Smith, ``{The information capacity of amplitude- and variance-constrained
  scalar Gaussian channels},'' \emph{Inform. Control}, vol.~18, pp. 203--219,
  1971.

\bibitem{Shamai_BarDavid_1995}
S.~Shamai and I.~Bar-David, ``The capacity of average and peak-power-limited
  quadrature gaussian channels,'' \emph{IEEE Trans. Inf. Theory}, vol.~41,
  no.~4, pp. 1060--1071, Jul. 1995.

\bibitem{AbouFaycal_Trott_Shamai_2001}
I.~C. Abou-Faycal, M.~D. Trott, and S.~Shamai, ``The capacity of discrete-time
  memoryless rayleigh-fading channels,'' \emph{IEEE Transactions on Information
  Theory}, vol.~47, no.~4, pp. 1290--1301, May 2001.

\bibitem{Tchamkerten_2004}
A.~Tchamkerten, ``On the discreteness of capacity-achieving distributions,''
  \emph{IEEE Trans. Inf. Theory}, vol.~50, no.~11, pp. 2773--2778, Nov. 2004.

\bibitem{Fahs_AbouFaycal_2012}
J.~J. Fahs and I.~C. Abou-Faycal, ``Using hermite bases in studying
  capacity-achieving distributions over awgn channels,'' \emph{IEEE Trans. Inf.
  Theory}, vol.~58, no.~8, pp. 5302--5322, Aug. 2012.

\bibitem{desurvire_2009}
E.~Desurvire, \emph{Classical and Quantum Information Theory: An Introduction
  for the Telecom Scientist}.\hskip 1em plus 0.5em minus 0.4em\relax Cambridge:
  Cambridge University Press, 2009.

\bibitem{Lomnitz_Feder}
Y.~Lomnitz and M.~Feder, ``Communication over individual channels,'' \emph{IEEE
  Trans. Inf. Theory}, vol.~57, no.~11, pp. 7333--7358, Nov. 2011.

\bibitem{Mitra_Stark_2001}
P.~P. Mitra and J.~B. Stark, ``Nonlinear limits to the information capacity of
  optical fibre communications,'' \emph{Nature}, vol. 411, pp. 1027--1030, Jan.
  2001.

\bibitem{Varshney_2008}
L.~R. Varshney, ``Transporting information and energy simultaneously,'' in
  \emph{IEEE Int. Sym. Inf. Theory}, Jul. 2008, pp. 1612--1616.

\bibitem{Essiambre_Kramer_Winzer_Foschini_Goebel}
R.~J. Essiambre, G.~Kramer, P.~J. Winzer, G.~J. Foschini, and B.~Goebel,
  ``Capacity limits of optical fiber networks,'' \emph{Journal of Lightwave
  Technology}, vol.~28, no.~4, pp. 662--701, Feb. 2010.

\bibitem{gradshteyn2007}
I.~S. Gradshteyn and I.~M. Ryzhik, \emph{Table of integrals, series, and
  products}, 7th~ed.\hskip 1em plus 0.5em minus 0.4em\relax Elsevier/Academic
  Press, Amsterdam, 2007.

\bibitem{Gallager_book}
R.~G. Gallager, \emph{Information Theory and Reliable Communication}.\hskip 1em
  plus 0.5em minus 0.4em\relax New York, NY, USA: John Wiley \& Sons, Inc.,
  1968.

\bibitem{Morsi_Jamali}
R.~Morsi, V.~Jamali, D.~W.~K. Ng, and R.~Schober, ``On the capacity of swipt
  systems with a nonlinear energy harvesting circuit,'' in \emph{IEEE Int' l
  Conference on Communications (ICC)}, May 2018, pp. 1--7.

\bibitem{Bayguzina_Clerckx_1802.06512}
E.~Bayguzina and B.~Clerckx, ``Modulation design for wireless information and
  power transfer with nonlinear energy harvester modeling,'' in \emph{IEEE 19th
  Int'l Workshop on Signal Processing Advances in Wireless Communications
  (SPAWC)}, Jun. 2018, pp. 1--5.

\bibitem{Rassouli_Clerckx_16}
B.~Rassouli and B.~Clerckx, ``On the capacity of vector gaussian channels with
  bounded inputs,'' \emph{IEEE Trans. Inf. Theory}, vol.~62, no.~12, pp.
  6884--6903, Dec. 2016.

\bibitem{Rassouli_Varasteh_Gunduz_17}
\BIBentryALTinterwordspacing
B.~Rassouli, M.~Varasteh, and D.~Gündüz, ``Gaussian multiple access channels
  with one-bit quantizer at the receiver,'' \emph{Entropy}, vol.~20, no.~9,
  2018. [Online]. Available: \url{http://www.mdpi.com/1099-4300/20/9/686}
\BIBentrySTDinterwordspacing

\bibitem{Agrell_Alvarado_Durisi_Karlsson}
E.~Agrell, A.~Alvarado, G.~Durisi, and M.~Karlsson, ``Capacity of a nonlinear
  optical channel with finite memory,'' \emph{Journal of Lightwave Technology},
  vol.~32, no.~16, pp. 2862--2876, Aug. 2014.

\bibitem{Huang_Meyn}
J.~Huang and S.~P. Meyn, ``Characterization and computation of optimal
  distributions for channel coding,'' \emph{IEEE Trans. Inf. Theory}, vol.~51,
  no.~7, pp. 2336--2351, Jul. 2005.

\bibitem{elGamal:book}
A.~E. Gamal and Y.-H. Kim, \emph{Network Information Theory}.\hskip 1em plus
  0.5em minus 0.4em\relax Cambridge University Press, 2011.

\bibitem{Hardy}
\BIBentryALTinterwordspacing
G.~H. Hardy, ``A theorem concerning fourier transforms,'' \emph{Journal of the
  London Mathematical Society}, vol. s1-8, no.~3, pp. 227--231, 1933. [Online].
  Available: \url{http://dx.doi.org/10.1112/jlms/s1-8.3.227}
\BIBentrySTDinterwordspacing

\bibitem{Luenberger_1969}
D.~Luenberger, \emph{Optimization by Vector Space Methods}.\hskip 1em plus
  0.5em minus 0.4em\relax New York: John Wiley \& Sons, Inc, 1969.

\bibitem{Laforgia_2010}
A.~Laforgia and P.~Natalini, ``Some inequalities for modified bessel
  functions,'' \emph{Journal of Inequalities and Applications}, vol. 2010,
  no.~1, p. 253035, Jan. 2010.

\end{thebibliography}

\begin{IEEEbiographynophoto}
{Morteza Varasteh}
received the MSc degree in Communications Systems Engineering from Sharif University of Technology, Tehran, Iran, in 2011. He received his PhD degree in Communications from Imperial College London, London, UK in 2016. He was a research associate in Imperial college from 2017 to 2019 working on the EPSRC-funded project "Simultaneous Information-Power Transfer". He is currently a lecturer (assistant professor) in the Computer Science and Electrical Engineering (CSEE) Department, University of Essex, Colchester Campus, UK. His research interests lie in the general area of information theory, wireless communications, optimization theory and machine learning.
\end{IEEEbiographynophoto}

\begin{IEEEbiographynophoto}
{Borzoo Rassouli}
received the M.Sc. degree in electrical engineering from university of Tehran, Iran in 2012, and the Ph.D. degree in communications engineering from Imperial College London, UK in 2016. He was a postdoctoral research associate at Imperial College from 2016 to 2018. In August 2018, he joined university of Essex as a lecturer (Assistant Professor). His research interests lie in the general areas of information theory and statistics.
\end{IEEEbiographynophoto}

\begin{IEEEbiographynophoto}
{Bruno Clerckx} [SM'17] received the M.S. and Ph.D. degrees in applied science from the Université Catholique de Louvain, Louvain-la-Neuve, Belgium, in 2000 and 2005, respectively. From 2006 to 2011, he was with Samsung Electronics, Suwon, South Korea, where he actively contributed to 3GPP LTE/LTE-A and IEEE 802.16m and acted as the Rapporteur for the 3GPP Coordinated Multi-Point (CoMP) Study Item. From 2014 to 2016, he was an Associate Professor with Korea University, Seoul, South Korea. He also held visiting research appointments at Stanford University, EURECOM, National University of Singapore, The University of Hong Kong and Princeton University. Since 2011, he has been with Imperial College London, first as a Lecturer from 2011 to 2015, then as a Senior Lecturer from 2015 to 2017, and now as a Reader. He is currently a Reader (Associate Professor) with the Electrical and Electronic Engineering Department, Imperial College London, London, U.K.

He has authored two books, 160 peer-reviewed international research papers, and 150 standards contributions, and is the inventor of 75 issued or pending patents among which 15 have been adopted in the specifications of 4G (3GPP LTE/LTE-A and IEEE 802.16m) standards. His research area is communication theory and signal processing for wireless networks. He has been a TPC member, a symposium chair, or a TPC chair of many symposia on communication theory, signal processing for communication and wireless communication for several leading international IEEE conferences. He is an Elected Member of the IEEE Signal Processing Society SPCOM Technical Committee. He served as an Editor for the IEEE TRANSACTIONS ON COMMUNICATIONS from 2011 to 2015 and the IEEE TRANSACTIONS ON WIRELESS COMMUNICATIONS from 2014 to 2018, and is currently an Editor for the IEEE TRANSACTIONS ON SIGNAL PROCESSING. He has also been a (lead) guest editor for special issues of the EURASIP Journal on Wireless Communications and Networking, IEEE ACCESS and the IEEE JOURNAL ON SELECTED AREAS IN COMMUNICATIONS. He was an Editor for the 3GPP LTE-Advanced Standard Technical Report on CoMP.

\end{IEEEbiographynophoto}

\end{document}